\documentstyle[12pt,a4wide,epsfig,amssymb]{article}

\def\lc{\Lambda_c}

\def\mk{m_K}
\def\mp{m_\pi}
\def\me{m_\eta}
\def\mep{m_{\eta'}}

\def\T{\textstyle}

\def\SS{\scriptscriptstyle}

\title{
\vspace*{-1.0cm}
\begin{flushright}
{\normalsize DO--TH 99/10\\LNF-99/016(P)\\[-12pt]
hep-ph/9906434}
\end{flushright}
\vspace{1.3cm}
{\Large \bf Analysis of \boldmath $\varepsilon'/\varepsilon$ in the
$1/N_c$ Expansion \unboldmath} 
\vspace*{0.8cm}
}
\author{ 
T.~Hambye$^{a\,}$\footnote{\small E-mail: hambye@lnf.infn.it}~, 
G.O.~K\"ohler$^{b\,}$\footnote{\small E-mail: 
koehler@doom.physik.uni-dortmund.de}~, 
E.A.~Paschos$^{b\,}$\footnote{\small E-mail: 
paschos@hal1.physik.uni-dortmund.de}~, and P.H. Soldan$^{b\,}
$\footnote{\small E-mail: soldan@doom.physik.uni-dortmund.de}\\[0.5cm]
\normalsize $a$: {\it INFN - Laboratori Nazionali di Frascati,
P.O. Box 13, I-00044 Frascati, Italy}\\[1mm] 
\normalsize $b$: {\it Institut f\"ur Physik, Universit\"at Dortmund 
D-44221 Dortmund, Germany}\\[3cm] 
}
\date{}
\begin{document}
\maketitle
\thispagestyle{empty}
\vspace*{-1.8cm}
\begin{abstract}
We present a new analysis of the ratio $\varepsilon'/\varepsilon$ which 
measures the direct CP violation in $K\rightarrow\pi\pi$ decays. We use the 
$1/N_c$ expansion within the framework of the effective chiral lagrangian for 
pseudoscalar mesons. The $1/N_c$ corrections to the hadronic matrix elements 
of all operators are calculated at leading order in the chiral expansion. 
Performing a simple scanning of the input parameters we obtain $1.5\cdot 
10^{-4}\leq \varepsilon'/\varepsilon\leq 31.6\cdot 10^{-4}$. We also 
investigate, in the chiral limit, the $1/N_c$ corrections to the operator 
$Q_6$ at next-to-leading order in the chiral expansion. We find large 
positive corrections which further enhance $\varepsilon'/\varepsilon$ and 
can bring the standard model prediction close to the measured value for 
central values of the parameters. Our result indicates that at the level
of the $1/N_c$ corrections a $\Delta I=1/2$ enhancement is operative for 
$Q_6$ similar to the one of $Q_1$ and $Q_2$ which dominate the CP 
conserving amplitude.
\end{abstract}
\vspace*{\fill}
%
\newpage
%
\section{Introduction}
There are two types of CP violation which appear in the neutral kaon 
system: direct and indirect. Direct CP violation occurs in the amplitudes
and will be the subject of this paper. Indirect violation occurs in the
physical states and is characterized in a phase convention independent 
way by the parameter $\varepsilon$. Indirect CP violation has been observed 
and is incorporated in the standard model, as a restriction to the CKM phase. 
Direct CP violation is described by the parameter $\varepsilon'$ whose
predictions require greater attention and has been the subject of several
investigations. The superweak theory \cite{wolf} predicts $\varepsilon'/
\varepsilon$ to be exactly zero. In the standard model the predictions for 
the ratio cover a wide range of values. Until recently, the experimental 
evidence for $\varepsilon'/\varepsilon$ was inconclusive. While the value 
$\mbox{Re}\,(\varepsilon'/\varepsilon)=(23\pm 7)\cdot10^{-4}$ reported by 
the NA31 collaboration at CERN \cite{barr} indicated direct CP violation, 
the result of the E731 collaboration at Fermilab \cite{gibb}, $\,(7.4\pm 
5.9)\cdot 10^{-4}$, was still compatible with a vanishing value. The new 
measurement of the KTeV collaboration~\cite{ktev},
\begin{equation}
\mbox{Re}\,(\varepsilon'/\varepsilon)\,\,=\,\,(28.0\pm 4.1)\cdot 10^{-4}\,, 
\end{equation}
is in agreement with the CERN experiment NA31 and rules out the 
superweak models. Additional information will be provided in the near 
future by the NA48 collaboration and by the KLOE experiment at DA$\Phi$NE. 
In view of the new experimental result, whose statistical uncertainty will 
be further reduced in the future, it is particularly interesting to 
investigate whether the quoted range and the weighted average can be 
accommodated in the standard model. 

Direct CP violation measures the relative phases of the decay amplitudes for
\begin{displaymath}
K^0\to \pi^0\pi^0\quad\quad {\rm{and}} \quad\quad K^0\to\pi^+\pi^-.
\end{displaymath}
The two pions in these decays can be in two isospin states, $I=0$ 
($\Delta I=1/2$) and $I=2$ ($\Delta I=3/2$). The two amplitudes acquire 
phases through final state strong interactions and also through the 
couplings of weak interactions. We can use Watson's theorem \cite{wat} 
to write them as
\vspace*{0.2em}
\noindent
\begin{equation}
\langle\pi\pi,I|{\cal H}_W|K^0\rangle = A_I e^{i\delta_I}\,,
\label{ampdef1}
\end{equation}
\begin{equation}
\langle\pi\pi,I|{\cal H}_W|\bar{K}^0\rangle = A_I^* e^{i\delta_I}\,,
\label{ampdef2}
\end{equation}
with $\delta_I$ being a phase of strong origin which is extracted
from $\pi-\pi$ scattering. The remaining amplitude $A_I$ contains
a phase of weak origin. Throughout the paper we use the following 
isospin decomposition:
\begin{equation}
A(K^0\rightarrow\pi^+\pi^-)\,=\,\sqrt{\frac{2}{3}}A_0e^{i\delta_0}
+\frac{1}{\sqrt{3}}A_2e^{i\delta_2}\,,\label{deco1}
\end{equation}
\begin{equation}
A(K^0\rightarrow\pi^0\pi^0)\,=\,\sqrt{\frac{2}{3}}A_0e^{i\delta_0}
-\frac{2}{\sqrt{3}}A_2e^{i\delta_2}\,,\label{deco2}
\end{equation}
\begin{equation}
A(K^+\rightarrow\pi^+\pi^0)\,=\,\sqrt{\frac{3}{2}}A_2e^{i\delta_2}\,.
\label{deco3}
\end{equation}

The parameter of direct CP violation can be written as
\begin{equation}
\frac{\varepsilon'}{\varepsilon}\,=\,\frac{\omega}{\sqrt{2}\,|\varepsilon|}
\,\left(\,\frac{\mbox{Im}A_2}{\mbox{Re}A_2}\,-\,\frac{\mbox{Im}A_0}
{\mbox{Re}A_0}\,\right)\,,\label{epspdef}
\end{equation}
with $\omega=\mbox{Re}A_2/\mbox{Re}A_0=1/22.2$. In Eq.~(\ref{epspdef}) 
we used the fact that, to a large degree of accuracy, the strong interaction 
phases of $\varepsilon'$ and $\varepsilon$ cancel in the ratio (see e.g.
Ref.~\cite{wins}). In order to obtain the numerical value of $\varepsilon'/
\varepsilon$ it is now necessary to calculate the two amplitudes 
($\mbox{Im}A_0$ and $\mbox{Im}A_2$) including their weak phases.

Using the operator product expansion, the $K\rightarrow\pi\pi$ amplitudes 
are obtained from the effective low-energy hamiltonian for $|\Delta S|=1$ 
transitions \cite{GLAM,VSZ,GWGP},
\begin{equation}
{\cal H}_{ef\hspace{-0.5mm}f}^{\SS \Delta S=1}=\frac{G_F}{\sqrt{2}}
\;\lambda_u\sum_{i=1}^8 c_i(\mu)\,Q_i(\mu)\hspace{1cm}(\mu < m_c)\,,
\label{ham}
\end{equation}
\begin{equation}
c_i(\mu)=z_i(\mu)+\tau y_i(\mu)\;,\hspace*{1cm}\tau=-\lambda_t/\lambda_u\;,
\hspace*{1cm}\lambda_q=V_{qs}^*\,V_{qd}^{}\;.\label{cidef}
\end{equation}
The arbitrary renormalization scale $\mu$ separates short- and long-distance 
contributions to the decay amplitudes. The Wilson coefficient functions 
$c_i(\mu)$ contain all the information on heavy-mass scales. The terms with 
the $z_i$'s contribute to the real parts of the amplitudes $A_0$ and $A_2$.  
The $y_i$'s, on the other hand, contribute to the imaginary parts and are
relevant for CP violating processes. The coefficient functions can be 
calculated for a scale $\mu \gtrsim 1\,$GeV using perturbative renormalization
group techniques. They were computed in an extensive next-to-leading logarithm
analysis by two groups \cite{BJM,CFMR}. The Wilson coefficients depend on the 
CKM elements; the $y_i$'s are multiplied by $\lambda_t$ which introduces CP
violation in the amplitudes. Finally, the calculation of the decays depends 
on the hadronic matrix elements of the local four-quark operators
\begin{equation}
\langle Q_i(\mu)\rangle_I \,\equiv\,\langle\pi\pi,\,I|\,Q_i(\mu)\,
|K^0\rangle\,,
\end{equation}
which constitute the non-perturbative part of the calculation. This is the 
main subject of this paper. The hadronic matrix elements will be calculated
using the $1/N_c$ expansion within the framework of the effective chiral
lagrangian for pseudoscalar mesons \cite{BBG2,BBG3,HKPSB}. In a previous 
article \cite{HKPSB} we already reported the results of ${\cal O}(p^0/N_c)$ 
for the operators $Q_6$ and $Q_8$. In this article we investigate one-loop 
corrections for all matrix elements relevant for~$\varepsilon'/\varepsilon$.

The local four-quark operators $Q_i(\mu)$ can be written, after Fierz 
reordering, in terms of color singlet quark bilinears:
\begin{eqnarray}
Q_1 &=& 4\,\bar{s}_L\gamma^\mu d_L\,\,\bar{u}_L\gamma_\mu u_L\,, 
\hspace*{2.49cm} 
Q_2 \,\,\,=\,\,\,\,4\,\bar{s}_L\gamma^\mu u_L\,\,\bar{u}_L
\gamma_\mu d_L\,, \label{qia} \\[4mm] 
Q_3 &=& \,4\,\sum_q \bar{s}_L\gamma^\mu d_L\,\,\bar{q}_L\gamma_\mu q_L\,,  
\hspace*{1.76cm}
Q_4 \,\,\,=\,\,\, \,4\,\sum_q \bar{s}_L\gamma^\mu q_L\,\,\bar{q}_L
\gamma_\mu d_L\,,\\[2mm]
Q_5 &=& \,4\,\sum_q \bar{s}_L\gamma^\mu d_L\,\,\bar{q}_R\gamma_\mu q_R\,, 
\hspace*{1.74cm}
Q_6 \,\,\,=\,\,\, \,-8\,\sum_q \bar{s}_L q_R\,\,\bar{q}_R d_L\,,\\[1mm]
Q_7 &=& \,4\,\sum_q \frac{3}{2}e_q\,\bar{s}_L\gamma^\mu d_L\,\,
\bar{q}_R \gamma_\mu q_R\,,\hspace*{1.0cm}
Q_8 \,\,\,=\,\,\, \,-8\,\sum_q \frac{3}{2}e_q\,\bar{s}_L q_R\,\,
\bar{q}_R d_L\,, \label{qio}
\end{eqnarray}
where the sum goes over the light flavors ($q=u,d,s$) and
\begin{equation}
q_{R,L}=\frac{1}{2}(1\pm\gamma_5)q\,,\hspace{1cm} 
e_q\,= (2/3,\,-1/3,\,-1/3)\,.
\end{equation}
$Q_3\,\,$-$\,\,Q_6$ arise from QCD penguin diagrams involving a virtual $W$ 
and a $c$ or $t$ quark, with gluons connecting the virtual heavy quark to 
light quarks. They transform as $(8_L,1_R)$ under $SU(3)_L\times SU(3)_R$ 
and contribute, in the isospin limit, only to $\Delta I=1/2$ transitions. 
$Q_7$ and $Q_8$ are electroweak penguin operators \cite{BW,BG1}. 

The imaginary parts of the amplitudes occurring in Eq.~(\ref{epspdef}) 
are those produced by the weak interaction. Thus we obtain the amplitudes
\begin{equation}
{\mbox Im}A_I\,\,=\,\,-\frac{G_F}{\sqrt{2}}\,\,\mbox{Im}\lambda_t\,\,
\Big|\sum_i y_i(\mu)\,\langle Q_i\rangle_I\Big|\,.
\end{equation}
Since the phase originating from the strong interactions is already 
extracted in Eq.~(\ref{ampdef1}), absolute values for the $\sum_i 
\,y_i\,\langle Q_i\rangle_I$ should be taken. We shall return to 
this point later~on.  

Collecting all terms together we arrive at the general expression
\begin{equation}
\frac{\varepsilon'}{\varepsilon}\,=\,\frac{G_F}{2}\frac{\omega}
{|\varepsilon|\,\mbox{Re}A_0}\,\mbox{ Im}\lambda_t\, 
\left[\,\Pi_0\,-\,\frac{1}{\omega}\Pi_2\,\right]\,,\label{epspsm}
\end{equation}
with
\begin{eqnarray}
\Pi_0&=&\Big|\sum_i\,y_i(\mu)\,\langle Q_i\rangle_0\Big|\,
\Big(1-\Omega_{\eta+\eta'}\Big)\,,\label{Pi0}\\[1mm]
\Pi_2&=&\Big|\sum_i\,y_i(\mu)\,\langle Q_i\rangle_2\Big|\,,\label{Pi2}
\end{eqnarray}
where $\Omega_{\eta+\eta'}\sim 0.25\pm 0.10$ takes into account the effect of 
the isospin breaking in the quark masses ($m_u\neq m_d$) \cite{BG1,don,lus}. 
We have written Eq.~(\ref{epspsm}) as a product of factors in order to 
emphasize the importance and uncertainty associated with each of them. 
The first factor contains known parameters and takes the numerical value 
$G_F\,\omega/(2\,|\varepsilon|\,\mbox{Re}A_0)\,=\,346\,\mbox{GeV}^{-3}$.  
The remaining terms are discussed in the following sections. Especially 
important to this analysis are the operators $Q_6$ and $Q_8$ which 
dominate the $I=0$ and $I=2$ contributions in Eqs.~(\ref{Pi0}) and 
(\ref{Pi2}), respectively. The terms $y_i\,\langle Q_i\rangle_2$ are 
enhanced by the factor $1/\omega$, and a crucial issue is whether the 
enhancement is strong enough to produce an almost complete cancellation 
with the $y_i\,\langle Q_i\rangle_0$ terms, leading to an approximately 
vanishing $\varepsilon'/\varepsilon$ even in the presence of direct CP 
violation, or whether the cancellation of the two terms is only moderate 
and a large value of $\varepsilon'/\varepsilon$ can be obtained within 
the standard model. 

The paper is organized as follows. In Section~\ref{CKM} we briefly recall 
the numerical values of the CKM elements relevant to this analysis. In 
Section~\ref{frame} we review the general framework of the effective 
low-energy calculation and discuss the matching of short- and long-distance 
contributions to the decay amplitudes. The next two sections contain our
results, which we present in two steps. As the chiral theory is an 
expansion in momenta, we keep the first two terms in the expansion and
calculate one-loop corrections to each term of the expansion separately.
Loop corrections to the lowest terms, in the momentum expansion, are
presented in Section~\ref{ana}; they are of ${\cal O}(p^0/N_c)$ for
the density operators and of ${\cal O}(p^2/N_c)$ for the current 
operators. In Section~\ref{hoc} we extend the one-loop corrections 
to the next order in momentum by calculating corrections to the
density operator $Q_6$ of ${\cal O}(p^2/N_c)$. Numerical results for
$\varepsilon'/\varepsilon$ are included in both of these sections.
Finally, our conclusions are contained in Section~\ref{con}.
%
%
\section{The CKM Elements \label{CKM}}

The second factor in Eq.~(\ref{epspsm}) originates from the CKM matrix. 
In the Wolfenstein para\-metri\-zation~\cite{wopar}
\vspace*{0.2em}
\noindent
\begin{equation}
\mbox{Im}\lambda_t\,=\,\mbox{Im}\,(V_{ts}^*\,V^{}_{td})\,=\, 
A^2\lambda^5\,\eta\,=\,V_{us}\, |V_{cb}|^2\,\eta\,,
\end{equation}
since $\lambda =V_{us}$ and $A=|V_{cb}|/\lambda^2$. Numerical values for 
the matrix elements are taken from the particle data group \cite{pdg}
and from Ref.~\cite{stoc}:
\begin{eqnarray}
|V_{us}| & = & 0.2196 \pm 0.0023\,,\\
|V_{cb}| & = & 0.040 \pm 0.002\,,\quad{\rm{and}}\\
|V_{ub}| & = & (3.56 \pm 0.56)\,\cdot\,10^{-3}\,.
\end{eqnarray}
The last parameter we need is the phase $\eta$ which is obtained from 
an analysis of the unitarity triangle whose overall scale is given by the 
value of $|V_{cb}|$. Such $\eta$ versus $\rho$ plots are now standard
\cite{ab98,china,ali,gnps,prs} and are obtained, primarily, from 
$|V_{ub}/V_{cb}|$ which produces a circular ring and from a hyperbola 
defined from the theoretical formula for $\varepsilon$. The position 
of the hyperbola depends on $m_t$, $|V_{cb}|$, and $\hat{B}_K$. The 
intersection of the two regions, together with constraints from the
observed $B_d^0-{\bar B}_d^0$ mixing parameterized by $\Delta M_d$ and 
the lower bound on $B_s^0-{\bar B}_s^0$ mixing, defines the physical 
ranges for $\eta$ and $\rho$. A very recent analysis can be found 
in Ref.~\cite{bosch}. The remaining theoretical uncertainties in 
this analysis are the values of the non-perturbative parameters 
$\hat{B}_K$ in $\varepsilon$, $F_{B_d}\sqrt{\hat{B}_d}$ in $(\Delta M)_d$, 
and $\xi=F_{B_s}\sqrt{\hat{B}_s}/F_{B_d}\sqrt{\hat{B}_d}$ in $(\Delta M)_d
/(\Delta M)_s$. $\hat{B}_K$ has been calculated by various methods which, 
unfortunately, give a large range of values. Two recent calculations are 
found in Refs.~\cite{BPdelta,hks}. Taking $\hat{B}_K=0.80 \pm 0.15$, 
$F_{B_d}\sqrt{\hat{B}_d}=200 \pm 40\,\mbox{MeV}$ \cite{bd,bdbs} and $\xi
=1.14 \pm 0.08$ \cite{bdbs,drapshar}, the authors of Ref.~\cite{bosch} 
obtain the following range for $\mbox{Im}\lambda_t$:
\vspace*{0.4em}
\begin{equation}
1.04\,\cdot\,10^{-4}\,\leq\,\mbox{Im}\lambda_t\,\leq\,1.63\,\cdot\,
10^{-4}\,, \label{ltval}
\end{equation}
where the experimentally measured values and the theoretical input 
parameters are scanned independently of each other, within the ranges 
given above. In Section~\ref{ana} we shall use this range in the 
numerical analysis of $\varepsilon'/\varepsilon$. 
%
%
\section{General Framework \label{frame}}
The method we use is the $1/N_c$ expansion introduced in 
Refs.~\cite{BBG2,BBG3}. In this approach, we expand the hadronic matrix
elements in powers of external momenta, $p$, and the ratio $1/N_c$. In an 
earlier article \cite{HKPSB} we investigated one-loop corrections to lowest 
order in the chiral expansion for the operators $Q_6$ and $Q_8$. The 
calculation of the one-loop corrections for current-current operators 
was done in Ref.~\cite{hks}, where predictions for the $\Delta I = 1/2$ 
rule were reported. 

To calculate the hadronic matrix elements we start from the effective chiral
lagrangian for pseudoscalar mesons which involves an expansion in momenta 
where terms up to ${\cal O}(p^4)$ are included \cite{GaL}. Keeping only 
(non-radiative) terms of ${\cal O}(p^4)$ which are leading in $N_c$,
for the lagrangian we obtain:
\begin{eqnarray}
{\cal L}_{ef\hspace{-0.5mm}f}&=&\frac{f^2}{4}\Big(
\langle D_\mu U^\dagger D^{\mu}U\rangle
+\frac{\alpha}{4N_c}\langle \ln U^\dagger -\ln U\rangle^2 
+r\langle {\cal M} U^\dagger+U{\cal M}^\dagger\rangle\Big) \nonumber\\[1.5mm] 
&&
+ L_1\langle D_\mu U^\dagger D^\mu U\rangle ^2 + L_2\langle D_\mu U^\dagger
D_\nu U\rangle \langle D^\mu U^\dagger D^\nu U\rangle + L_3\langle D_\mu
U^\dagger D^\mu U D_\nu U^\dagger D^\nu U\rangle \hspace*{4mm} 
\nonumber \\[2.5mm]
&& 
+ rL_5\langle D_\mu U^\dagger D^\mu U({\cal M}^\dagger U+U^\dagger{\cal M})
\rangle + r^2L_8\langle {\cal M}^\dagger U{\cal M}^\dagger U+{\cal M} 
U^\dagger{\cal M} U^\dagger \rangle\nonumber\\[2.5mm]
&& 
+ r^2 H_2\langle{\cal M}^\dagger {\cal M}\rangle \label{lagr}\,,
\end{eqnarray}
with $D_\mu U=\partial_\mu U-ir_\mu U+iUl_\mu$, $\langle A\rangle$ denoting 
the trace of $A$ and ${\cal M}=\mbox{diag}( m_u,m_d,m_s)$. $l_\mu$~and~$r_\mu$
are left- and right-handed gauge fields, respectively, $f$ and $r$ are 
parameters related to the pion decay constant $F_\pi$ and to the quark 
condensate, with $r = - 2 \langle \bar{q}q\rangle/f^2$. The complex matrix 
$U$ is a non-linear representation of the pseudoscalar meson nonet:
\begin{equation}
U=\exp\frac{i}{f}\Pi\,,\hspace{1cm} \Pi=\pi^a\lambda_a\,,\hspace{1cm} 
\langle\lambda_a\lambda_b\rangle=2\delta_{ab}\,, 
\end{equation}
where, in terms of the physical states,
\begin{equation}
\Pi=\left(
\begin{array}{ccc}
\T\pi^0+\frac{1}{\sqrt{3}}a\eta+\sqrt{\frac{2}{3}}b\eta'
& \sqrt2\pi^+ & \sqrt2 K^+  \\[2mm]
\sqrt2 \pi^- & \T
-\pi^0+\frac{1}{\sqrt{3}}a\eta+\sqrt{\frac{2}{3}}b\eta' & \sqrt2 K^0 \\[2mm]
\sqrt2 K^- & \sqrt2 \bar{K}^0 & 
\T -\frac{2}{\sqrt{3}}b\eta+\sqrt{\frac{2}{3}}a\eta'
\end{array} \right)\,,
\end{equation}
with
\begin{equation}
a= \cos \theta-\sqrt{2}\sin\theta\,, \hspace{1cm}
\sqrt{2}b=\sin\theta+\sqrt{2}\cos\theta\,, \label{abth}
\end{equation}
The conventions and definitions we use are the same as those in
Refs.~\cite{HKPSB,hks}. In particular, we introduce the singlet $\eta_0$
in the same way and with the same value for the $U_A(1)$ symmetry breaking
parameter, $\alpha=\me^2+\mep^2-2\mk^2\simeq 0.72\,\mbox{GeV}^2$,
corresponding to the $\eta-\eta'$ mixing angle $\theta = -19^\circ$
\cite{eta}. The bosonic representations of the quark currents and 
densities are defined in terms of (functional) derivatives of the 
chiral action and the lagrangian, respectively:
\begin{eqnarray}
\bar{q}_{iL}\gamma^\mu q_{jL}&\hspace*{-1mm}\equiv&
\hspace*{-1mm}\frac{\delta S}{\delta(l_\mu(x))_{ij}}\,=\,
-i\frac{f^2}{2}\big(U^\dagger\partial^\mu U\big)_{ji} \nonumber\\[2.5mm]
&&\hspace*{-1mm} 
+irL_5\big(\partial^\mu U^\dagger{\cal M}-{\cal M}^\dagger\partial^\mu U
+\partial^\mu U^\dagger U{\cal M}^\dagger U-U^\dagger{\cal M} U^\dagger
\partial^\mu U\big)_{ji}\,,\label{curr}\\[2mm]
\bar{q}_{iR} q_{jL}
&\equiv&-\frac{\delta{\cal L}_{ef\hspace{-0.5mm}f}}{\delta{\cal M}_{ij}}\,
=-r\Big(\frac{f^2}{4}U^\dagger+L_5\partial_\mu U^\dagger\partial^\mu U 
U^\dagger +2rL_8U^\dagger{\cal M} U^\dagger+rH_2{\cal M}^\dagger\Big)_{ji}
\;,\hspace*{7mm}\label{dens}
\end{eqnarray}
and the right-handed currents and densities are obtained by parity 
transformation. Eqs. (\ref{curr}) and~(\ref{dens}) allow us to express 
the four-fermion operators in terms of the pseudoscalar meson fields.
The low-energy couplings $L_1$, $L_2$, and $\,L_3$ do not occur in 
the mesonic densities in Eq.~(\ref{dens}). Furthermore, at tree level
they do not contribute to the matrix elements of the current-current 
operators and have been omitted in Eq.~(\ref{curr}). It is now 
straightforward to calculate the tree level (leading-$N_c$) matrix 
elements from the mesonic form of the 4-quark operators.

For the $1/N_c$ corrections to the matrix elements $\langle Q_i\rangle_I$ 
we calculated chiral loops as described in Refs.~\cite{HKPSB,hks}. The
factorizable contributions, on the one hand, refer to the strong sector of
the theory and give corrections whose scale dependence is absorbed in the 
renormalization of the effective chiral lagrangian. This property is
obvious in the case of the (conserved) currents and was demonstrated 
explicitly in the case of the bosonized densities \cite{HKPSB,TP}. 
Consequently, the factorizable loop corrections can be computed within 
dimensional regularization. The non-factorizable corrections, on the other 
hand, are UV divergent and must be matched to the short-distance part. They 
are regularized by a finite cutoff which is identified with the short-distance
renormalization scale \cite{BBG3,HKPSB,china,hks,DO1}. The definition of the 
momenta in the loop diagrams, which are not momentum translation invariant, 
was discussed in detail in Refs.~\cite{HKPSB,TP}. A consistent matching is 
obtained by considering the two currents or densities to be connected to 
each other through the exchange of a color singlet boson and by assigning 
the same momentum to it in the long- and short-distance 
regions~\cite{HKPSB,BB,BGK,FG,PSOrsay,egpt}.

For the short-distance coefficient functions, we use both the 
leading logarithmic (LO) and the next-to-leading logarithmic (NLO) 
values.\footnote{We treat the Wilson coefficients as leading order in 
$1/N_c$ since the large logarithms arising from the long renormalization 
group evolution from $(m_t,M_W)$ to $\mu\simeq {\cal O}(1\,\mbox{GeV})$ 
compensate for the $1/N_c$ suppression.} The published values for the 
Wilson coefficients are tabulated for scales equal to or larger than 
$1\,\mbox{GeV}$ \cite{BJM,CFMR,BBL}. In Appendix~\ref{append} we present 
tables for the coefficient functions at scales $0.6\leq\mu\leq 1.0\,
\mbox{GeV}$ calculated with the same analytic formulas and communicated 
to us by M.~Jamin \cite{jam}. The NLO values are scheme dependent and 
are calculated within naive dimensional regularization (NDR) and in 
the 't Hooft-Veltman scheme (HV), respectively. The absence of any 
reference to the renormalization scheme dependence in the effective 
low-energy calculation, at this stage, prevents a complete matching at 
the next-to-leading order~\cite{ab98}. Nevertheless, a comparison of the 
numerical results obtained from the LO and NLO coefficients is useful 
in order to estimate the uncertainties associated with it and to test
the validity of perturbation theory.
%
%
\section{Analysis of \boldmath $\varepsilon'/\varepsilon$ 
\unboldmath \label{ana}}
In the twofold expansion in powers of external momenta and in $1/N_c$ we 
must investigate, at next-to-leading order, the tree level contributions 
from the ${\cal O}(p^2)$ and the ${\cal O}(p^4)$ lagrangian, and the 
one-loop contribution from the ${\cal O}(p^2)$ lagrangian, that is to say, 
the $1/N_c$ corrections at lowest order in the chiral expansion. In this 
section we combine our results and report values for $\varepsilon'/
\varepsilon$ up to these orders.
\subsection{Long-Distance Contributions \label{hme1}}
The hadronic matrix elements for all the operators are calculated following
the method described in the previous section. As mentioned above, we consider 
the bilinear quark operators to be connected to each other through the 
exchange of a color singlet boson, whose momentum is chosen to be the 
variable of integration. This is our matching procedure described in 
Ref.~\cite{HKPSB}. For current-current operators, the tree level 
contributions from the ${\cal O}(p^2)$ and ${\cal O}(p^4)$ lagrangian 
and the one-loop contribution from the ${\cal O}(p^2)$ lagrangian are 
${\cal O}(p^2)$, ${\cal O}(p^4)$, and ${\cal O}(p^2/N_c)$, respectively. 
For density-density operators they are ${\cal O}(p^0)$, ${\cal O}(p^2)$, 
and ${\cal O}(p^0/N_c)$, respectively. The numerical results for the matrix 
elements to these orders are given in Tabs.~\ref{tab1} and~\ref{tab2} as a 
function of the cutoff scale. These values were obtained in 
Refs.~\cite{HKPSB,hks} using the following values for the various 
parameters~\cite{pdg}:
\[
\begin{array}{lclcllcl}
m_\pi &\equiv& \big(m_{\pi^0}+m_{\pi^+}\big)/2&=& 137.3\,\,\mbox{MeV}\,,
\hspace{0.6cm}&F_\pi&=&92.4\,\,\mbox{MeV}\,,\\[1.1mm]
\mk   &\equiv& \big(m_{K^0}+m_{K^+}\big)/2 &=& 495.7\,\,\mbox{MeV}\,,
&F_K&=&113\,\,\,\mbox{MeV}\,,\\[1.1mm]
m_\eta   &=& 547.5\,\,\mbox{MeV}\,,&&&\theta&=&-19^\circ\,,
\\[1.1mm]
m_{\eta'}&=& 957.8\,\,\mbox{MeV}\,.&&& & &\,\,
\mbox\,\\[1.1mm]
\end{array}
\]
Note that the matrix elements generally contain a non-vanishing imaginary 
part (cutoff independent at the one-loop level) which comes from the 
on-shell ($\pi-\pi$) rescattering. 
\begin{table}[t]
\begin{eqnarray*}
\begin{array}{|c||c|c|c|c|c|c|}\hline
\lc&0.6\,\,\mbox{GeV}&0.7\,\,\mbox{GeV}&0.8\,\,\mbox{GeV}&0.9\,\,
\mbox{GeV}&1.0\,\,\mbox{GeV}&\,\,\mbox{Im}\langle Q_{i}\rangle_{0}\,\,\\ 
\hline\hline
\rule{0cm}{5mm}
\langle Q_{1} \rangle_{0} & -33.2 & -40.2 & -48.2 & -57.3 & -67.4  &  -5.55 i \\[0.5mm]
\langle Q_{2} \rangle_{0} & 58.8 & 68.8 & 79.9 & 92.4 & 106  & 11.1 i \\[0.5mm]
\langle Q_{3} \rangle_{0} & 0.05 & 0.03 & -0.02 & -0.12 & -0.26  & 0 \\[0.5mm]
\langle Q_{4} \rangle_{0} & 92.1 & 109 & 128 & 150 & 173  & 16.6 i \\[0.5mm]
\langle Q_{5} \rangle_{0} & -0.05 & -0.03 & 0.02 & 0.12 & 0.26  & 0 \\[0.5mm]
\langle Q_{6} \rangle_{0} & -38.6 & -33.7 & -29.4 & -25.5 & -21.9  & 0 \\[0.5mm]
\langle Q_{7} \rangle_{0} & 40.1 & 46.6 & 54.1 & 62.6 & 72.2  & 8.32 i \\[0.5mm]
\langle Q_{8} \rangle_{0} & 119 & 119 & 119 & 118 & 117 & 36.7 i \\[0.5mm]
\hline
\end{array}
\end{eqnarray*}
\caption{Real and imaginary parts (last column) for the hadronic matrix 
elements of $Q_{1,\ldots,5,7}$ (in units of $10^{6}\cdot\mbox{MeV}^{3}$) 
and $Q_{6,8}$ [\,in units of $R^2 \cdot \mbox{MeV}$, with $R\equiv 
2m_K^2/(m_s+m_d)\,$]. The values are for the $I = 0$ amplitudes in the 
isospin limit ($m_u=m_d$) and for various values of the cutoff $\lc$.
\label{tab1}}
\vspace*{-0.1cm}
\end{table}
\begin{table}[t]
\begin{eqnarray*}
\begin{array}{|c||c|c|c|c|c|c|}\hline
\lc&0.6\,\,\mbox{GeV}&0.7\,\,\mbox{GeV}&0.8\,\,\mbox{GeV}&0.9\,\,
\mbox{GeV}&1.0\,\,\mbox{GeV}& \,\,\mbox{Im}\langle Q_{i}\rangle_{2}\,\,\\ 
\hline\hline 
\rule{0cm}{5mm}
\langle Q_{1} \rangle_{2} & 2.51 & -2.26 & -7.77 & -14.0 & -21.1  & -3.45 i \\[0.5mm]
\langle Q_{2} \rangle_{2} & 2.51 & -2.26 & -7.77 & -14.0 & -21.1  & -3.45 i \\[0.5mm]
\langle Q_{7} \rangle_{2} & -10.7 & -6.27 & -1.15 & 4.67 & 11.2  &  5.18 i \\[0.5mm]
\langle Q_{8} \rangle_{2} & 35.3 & 31.2 & 27.2 & 23.2 &  18.8 & -11.5 i \\[0.5mm]
\hline
\end{array}
\end{eqnarray*}
\caption{
Same results as in Tab.~\ref{tab1} for the $I = 2$ amplitudes.\label{tab2}}
\end{table}

It is customary to parameterize the hadronic matrix elements in terms 
of the bag factors $B_{i}^{(1/2)}$ and $B_{i}^{(3/2)}$, which quantify 
the deviations from the values obtained in the vacuum saturation 
approximation~\cite{VSA}:
\begin{eqnarray}
B_{i}^{(1/2)}&=&\frac{\mbox{Re}\langle Q_{i} \rangle_{0}}
{\langle Q_{i} \rangle_{0}^{\scriptsize \mbox{VSA}}}\,,\quad i \, \in \, 
\{1,\ldots ,8\}\,,  \label{bag0}\\[2mm]
B_{i}^{(3/2)}&=&\frac{\mbox{Re}\langle Q_{i} \rangle_{2}}
{\langle Q_{i} \rangle_{2}^{\scriptsize \mbox{VSA}}}\,,\quad i \, \in \, 
\{1,2,7,8\} \,.\label{bag2}
\end{eqnarray}
{
\begin{table}[t]
\begin{eqnarray*}
\begin{array}{|c||c|c|c|c|c|}\hline
\lc&0.6\,\,\mbox{GeV}&0.7\,\,\mbox{GeV}&
0.8\,\,\mbox{GeV}&0.9\,\,\mbox{GeV}&1.0\,\,\mbox{GeV} \\ 
\hline\hline 
\rule{0cm}{5mm} 
B_{1}^{(1/2)}& 8.24 & 9.98 & 12.0 & 14.2 & 16.6  \\[0.5mm]
B_{2}^{(1/2)}& 2.91 & 3.41 & 3.96 & 4.57 &  5.23 \\[0.5mm]
B_{3}^{(1/2)}& 0.004 & 0.002 & -0.002 & -0.010 & -0.021 \\[0.5mm]
B_{4}^{(1/2)}& 2.54 & 3.00 & 3.53 & 4.13 & 4.75 \\[0.5mm]
B_{5}^{(1/2)}& 0.0009 & 0.0005 & -0.0003 & -0.0014 & 
-0.0020 \\[0.5mm]
B_{6}^{(1/2)}& 1.10 & 0.96 & 0.84 & 0.72 & 0.62 \\[0.5mm]
B_{7}^{(1/2)}& 0.16 & 0.18 & 0.21 & 0.23 & 0.26 \\[0.5mm]
B_{8}^{(1/2)}& 1.21 & 1.21 & 1.21 & 1.20 & 1.19 \\[0.5mm]
\hline
\end{array}
\end{eqnarray*}
\caption{Bag parameters for the $I = 0$ amplitudes, shown for various values 
of the cutoff. $B^{(1/2)}_{5,\,7,\,8}$ depend on $R\simeq 2m_K^2/m_s$ and 
are calculated for a running $m_s(\mu=\lc)$ at the leading logarithmic 
order ($\Lambda_{\mbox{\tiny QCD}}=325\,\mbox{MeV}$) with $m_s(1\,
\mbox{GeV})=175\,\mbox{MeV}$.\label{tab3}}
\vspace*{4mm}
\end{table}
\begin{table}[t]
\begin{eqnarray*}
\begin{array}{|c||c|c|c|c|c|c|}\hline
\lc&0.6\,\,\mbox{GeV}&0.7\,\,\mbox{GeV}&
0.8\,\,\mbox{GeV}&0.9\,\,\mbox{GeV}&1.0\,\,\mbox{GeV} \\ 
\hline\hline 
\rule{0cm}{5mm}
B_{1}^{(3/2)}& 0.11 & -0.10 & -0.34 & -0.61 & -0.92 \\[0.5mm]
B_{2}^{(3/2)}& 0.11 & -0.10 & -0.34 & -0.61 & -0.92 \\[0.5mm]
B_{7}^{(3/2)}& -0.10 & -0.06 & -0.01 & 0.04 & 0.09  \\[0.5mm]
B_{8}^{(3/2)}& 0.64 & 0.56 & 0.49 & 0.42 & 0.34  \\[0.5mm]
\hline
\end{array}
\end{eqnarray*}
\caption{Same results as in Tab.~\ref{tab3} for the $I = 2$ amplitudes. 
\label{tab4}}
\end{table}
}
The VSA expressions for the matrix elements are taken from 
Eqs.~(XIX.11)\,-\,(XIX.28) of Ref.~\cite{BBL}, and the corresponding 
numerical values can be found in Refs.~\cite{HKPSB,hks}. We list the bag 
parameters in Tabs.~\ref{tab3} and~\ref{tab4}.\footnote{The definition of 
the bag parameters in Eqs.~(\ref{bag0}) and~(\ref{bag2}) refers to the 
complete sum of the factorizable and non-factorizable terms in the hadronic
matrix elements. Therefore we are free of any of the infrared problems 
discussed in Ref.~\cite{BPdelta}, which occur for $Q_6$ with other 
definitions of $B_6^{(1/2)}$.} One might note that the values of the 
$B$~factors contain the real parts of the matrix elements and 
not their absolute values. For the amplitudes appearing in 
Eqs.~(\ref{epspsm})\,-\,(\ref{Pi2}) we need both real and imaginary
parts for the matrix elements. We calculated the imaginary parts in 
the $1/N_c$ expansion and included their values in Tabs.~\ref{tab1} 
and~\ref{tab2}. They are produced by the imaginary parts of the one-loop
diagrams, as required by unitarity. In order to study the sensitivity of
the results on the imaginary part we calculated the matrix elements by two 
methods (for a discussion of this point see Ref.~\cite{hks}). In the first 
method, we obtain the absolute values by correcting the real parts using 
the phenomenological phases. This procedure has also been followed in 
Refs.~\cite{BEF,BEFL}. In the second method we assume zero phases and 
use the real parts of the matrix elements. The second method, to a 
large degree of accuracy, corresponds to adopting the phases  
obtained in the $1/N_c$ expansion.  

Analytical formulas for all matrix elements are given in 
Refs.~\cite{HKPSB,hks}. Among them four are particularly 
interesting and important, and we repeat them here:
\begin{eqnarray}
\langle Q_1\rangle_0
&=& -\frac{1}{\sqrt{3}}F_\pi\left( \mk^2 - \mp^2 \right)
\left[1+\frac{4\hat{L}_5^r}{F_\pi^2}m_\pi^2+
\frac{1}{(4\pi)^2F_\pi^2}\right.
\nonumber\\[0.5mm]
&&\times\left.\left( 6\Lambda_c^2-\Big(\frac{1}{2}\mk^2
+6m_\pi^2\Big)\log\Lambda_c^2\,+\,\cdots\,
\right)\right]\,,\label{rm1}\\[2mm]
\langle Q_2\rangle_0
&=& \frac{2}{\sqrt{3}}F_\pi\left( \mk^2 - \mp^2 \right)
\left[1+\frac{4\hat{L}_5^r}{F_\pi^2}m_\pi^2+
\frac{1}{(4\pi)^2F_\pi^2}\right.
\nonumber \\[0.5mm]
&&\left.\times\left(\frac{15}{4}\Lambda_c^2+\Big(\frac{11}{8}\mk^2
-\frac{15}{4}m_\pi^2\Big)\log\Lambda_c^2\,+\,\cdots\,\right)\right]\,,
\label{rm2}\\[3mm]
\langle Q_6\rangle_0
&=&-\frac{4\sqrt{3}}{F_\pi}\left(\frac{2m_K^2}{m_s+m_d}\right)^2
(m_K^2-m_\pi^2) \left[\hat{L}_5^r-\frac{3}{16\,(4\pi)^2}\,\log\Lambda_c^2
\,+\,\cdots\,\right]\label{rm3}\,,\\[3mm]
\langle Q_8\rangle_2
&=&\frac{\sqrt{3}}{2\sqrt{2}}F_\pi\left(\frac{2m_K^2}{m_s+m_d}
\right)^2\left[1+\frac{8m_K^2}{F_\pi^2}\,\Big(\hat{L}_5^r-2\hat{L}_8^r\Big) 
-\frac{4m_\pi^2}{F_\pi^2}\,\Big(3\hat{L}_5^r-8\hat{L}_8^r\Big)\right.
\hspace*{4mm}\nonumber\\[1mm]
&& 
-\left.\frac{1}{(4\pi)^2F_\pi^2}\Big(m_K^2-m_\pi^2+\frac{2}{3}
\alpha\Big)\log\Lambda_c^2 \,+\,\cdots\,\right]\,,\label{rm4}
\end{eqnarray}
where the ellipses denote finite terms, which for brevity are not written 
explicitly here, but have been included in the numerical analysis (in 
particular, they provide the mass terms which make the logarithms 
dimensionless). The constants $\hat{L_5^r}$ and $\hat{L_8^r}$ are 
renormalized couplings and defined through the relations~\cite{HKPSB}
\begin{equation} 
\frac{F_K}{F_\pi}\,\equiv\,1+\frac{4\hat{L}_5^r}{F_\pi^2}(\mk^2-\mp^2)
\label{kp1}
\vspace*{2mm}
\end{equation}
and
\begin{equation} 
\frac{m_K^2}{m_\pi^2}\,\equiv\,\frac{\hat{m}+m_s}{2\hat{m}}
\left[1-\frac{8(m_K^2-m_\pi^2)}{F_\pi^2}(\hat{L}_5^r-2\hat{L}_8^r)\right]
\,.\label{kp3}
\end{equation}
Their values are $\hat{L}_5^r=2.07\cdot 10^{-3}$ and $\hat{L}_8^r=1.09
\cdot 10^{-3}$. $\hat{L}_8^r$ has a small dependence on the ratio 
$m_s/m_d$ and we shall use the (central) value (of) $m_s/m_d=24.4 
\pm 1.5$~\cite{HL}.

In Eqs.~(\ref{rm1})\,-\,(\ref{rm4}) we have summed the factorizable
contributions [\,first two terms on the r.h.s.~of Eqs.~(\ref{rm1}) and 
(\ref{rm2}), first term on the r.h.s.~of Eq.~(\ref{rm3}), and first 
three terms on the r.h.s.~of Eq.~(\ref{rm4})\,] and the non-factorizable 
contributions. Factorizable terms originate from tree level diagrams
or from one-loop corrections to a single current or density whose 
scale dependence is absorbed in the renormalization of the effective 
chiral lagrangian (i.e., in $F_\pi$, $F_K$, $\hat{L}_5^r$, $\hat{L}_8^r$, 
and in the renormalization of the masses and wave functions). Finite terms
from the factorizable loop diagrams for $\langle Q_6\rangle_0$ and    
$\langle Q_8\rangle_2$ are not absorbed completely and must be included 
in the numerical analysis~\cite{HKPSB}. We note that the couplings $L_1$, 
$L_2$, and $L_3$ do not contribute to the matrix elements of $Q_6$ and 
$Q_8$ to ${\cal O}(p^2)$ and to ${\cal O}(p^4)$ for the current-current 
operators. The non-factorizable contributions, on the other hand, are UV 
divergent and must be matched to the short-distance part. As we already 
discussed above, they are regularized by a finite cutoff, $\Lambda_c$, 
which is identified with the renormalization scale $\mu$ of QCD. 

We discuss next Eqs.~(\ref{rm1})\,-\,(\ref{rm4}) which have several 
interesting properties \cite{HKPSB,hks}. First, the VSA values for
$\langle Q_1\rangle_0$ and $\langle Q_2\rangle_0$ are far too small 
to account for the large $\Delta I=1/2$ enhancement observed in the CP 
conserving amplitudes. Using the large-$N_c$ limit [\,$B_1^{(1/2)}=3.05$, 
$B_2^{(1/2)}=1.22$\,] improves the agreement between theory and the 
experimental result, but it still provides a gross underestimate. However, 
the non-factorizable $1/N_c$ corrections in Eqs.~(\ref{rm1}) and~(\ref{rm2}) 
contain quadratically divergent terms which are not suppressed with respect 
to the tree level contribution, since they bring in a factor of $\Delta
\equiv\Lambda_c^2/(4\pi F_\pi)^2$ and have large prefactors which, in 
some cases, can be as large as six in Eq.~(\ref{rm1}). Quadratic terms 
in $\langle Q_1\rangle_0$ and $\langle Q_2\rangle_0$ produce a large 
enhancement (see Tab.~\ref{tab3}) which brings the $\Delta I=1/2$ 
amplitude in agreement with the observed value~\cite{hks}. Corrections 
beyond the chiral limit ($m_q=0$) in Eqs.~(\ref{rm1})\,-\,(\ref{rm2}) 
are suppressed by a factor of $\delta = m_{K,\pi}^2/(4\pi F_\pi)^2 
\simeq 20\,\%$ and were found to be numerically small. 

The case of $\langle Q_6\rangle_0$ and $\langle Q_8\rangle_2$ is different 
from that of $\langle Q_{1,2}\rangle_0$. The leading-$N_c$ values are very
close to the corresponding VSA values. Moreover, the non-factorizable loop 
corrections in Eqs.~(\ref{rm3}) and~(\ref{rm4}), which are of ${\cal O}
(p^0/N_c)$, are found to be only logarithmically divergent~\cite{HKPSB}. 
Consequently, in the case of $\langle Q_8\rangle_2$ they are suppressed by 
a factor of $\delta$ compared to the leading ${\cal O}(p^0)$ term and are 
expected to be of the order of $20\,\%$ to $50\,\%$ depending on the 
prefactors. We note that Eq.~(\ref{rm4}) is a full leading plus 
next-to-leading order analysis of the $Q_8$ matrix element. The case of 
$B_6^{(1/2)}$ is more complicated since the ${\cal O}(p^0)$ term vanishes 
for $Q_6$. Nevertheless, the non-factorizable loop corrections to this term 
remain and have to be matched to the short-distance part of the amplitudes. 
These ${\cal O}(p^0/N_c)$ non-factorizable corrections must be considered at 
the same level, in the twofold expansion, as the ${\cal O}(p^2)$ tree 
contribution. Consequently, a value of $B_6^{(1/2)}$ around one [\,which 
corresponds to the ${\cal O}(p^2)$ term alone\,] is not a priori expected. 
However, numerically it turns out that the ${\cal O}(p^0/N_c)$ contribution 
is only moderate (see Tab.~\ref{tab3}). This property can be understood from 
the $(U^\dagger)_{dq}(U)_{qs}$ structure of the $Q_6$ operator which vanishes 
to ${\cal O}(p^0)$ implying that the factorizable and non-factorizable 
${\cal O}(p^0/N_c)$ contributions cancel to a large extent~\cite{HKPSB}. 
The fact that the factorizable and non-factorizable terms to this order 
have infrared divergences which must cancel in the sum of both contributions 
gives another qualitative hint for a value of $B_6^{(1/2)}$ remaining around 
one~\cite{BPdelta}. This explains why for $Q_6$ to ${\cal O}(p^0/N_c)$ we do 
not observe a $\Delta I=1/2$ enhancement similar to the one for $Q_1$ and
$Q_2$ in the CP conserving amplitude. The leading-$N_c$ values for 
$B_6^{(1/2)}$ and $B_8^{(3/2)}$ are therefore more efficiently protected 
from possible large $1/N_c$ corrections of the ${\cal O}(p^2)$ lagrangian 
than $B_{1,2}^{(1/2)}$. The effect of the ${\cal O}(p^0/N_c)$ term is however 
important for $B_6^{(1/2)}$ as for $B_8^{(3/2)}$ because it gives rise, 
in general, to a noticeable dependence on the cutoff scale \cite{HKPSB}, 
which is relevant for the matching with the short-distance part (see 
below). We note that $B_8^{(3/2)}$ shows a scale dependence which is very 
similar to the one of $B_6^{(1/2)}$ (see Tabs.~\ref{tab3} and~\ref{tab4}) 
leading to a stable ratio $B_6^{(1/2)}/B_8^{(3/2)}$ over a large range of 
scales around the value $B_6^{(1/2)}/B_8^{(3/2)}\simeq 1.77 \pm 0.05$ where 
the error refers to the variation of $\Lambda_c$ between $600$\,MeV and 
$1$\,GeV. The ${\cal O}(p^0/N_c)$ corrections consequently make the 
cancellation of $Q_6$ and $Q_8$ in $\varepsilon'/\varepsilon$ less 
effective. 

Finally, $B_1^{(3/2)}$ and $B_2^{(3/2)}$ were found to be too small 
to account for the measured value of the $\Delta I=3/2$ amplitude for 
small values of the cutoff and even become negative for large values 
of $\Lambda_c$ (see Tab.~\ref{tab4}). Due to an almost complete 
cancellation of the two numerically leading terms~\cite{hks} $B_1^{(3/2)}$ 
and $B_2^{(3/2)}$ are expected to be sensitive to corrections from higher 
order terms and/or higher resonances.\footnote{As explained in 
Ref.~\cite{hks}, the sensitivity of $B_{1,2}^{(1/2)}$ to corrections 
from higher order terms is expected to be smaller. Therefore, the fact 
that the $1/N_c$ expansion, at this stage, does not reproduce the 
$\Delta I=3/2$ amplitude does not imply that the $\Delta I=1/2$ amplitude 
cannot be calculated to a sufficient degree of accuracy. This point was 
also illustrated in Ref.~\cite{BPdelta} where it was shown that higher 
order corrections investigated with a Nambu Jona-Lasinio model are much 
larger for the $\Delta I=3/2$ channel than for the $ \Delta I=1/2$ one.}
For this reason, even though the effect of $B_{1,2}^{(3/2)}$ is small, 
in the analysis of $\varepsilon'/\varepsilon$ we will not use the values 
listed in the table but we will extract the parameters $B_1^{(3/2)}$ and 
$B_2^{(3/2)}$ from the experimental value of $\mbox{Re}A_2$. This point 
is further discussed in the next section.

\renewcommand{\topfraction}{1.0}
\subsection{Numerical Results \label{numdis1}}
Collecting together the values of the matrix elements in Tabs.~1 and~2
and the values of the Wilson coefficients in Appendix~\ref{append} we can 
give now the numerical results for $\varepsilon'/\varepsilon$. As we already 
mentioned above, we use the real parts of the matrix elements and consider 
two cases. In the first case, we use the real part of our calculation and 
the phenomenologically determined values for the final state interaction 
phases, $\delta_0=(34.2\pm 2.2)^\circ$ and $\delta_2=(-6.9\pm 0.2)^\circ$
\cite{phases}, in order to arrive from Eqs.~(\ref{Pi0}) and~(\ref{Pi2}) to
\begin{eqnarray}
\Pi_0&=&\frac{1}{\cos\delta_0}\,\sum_i\,y_i(\mu)\,\mbox{Re}\langle 
Q_i\rangle_0\,(1-\Omega_{\eta+\eta'})\,,\label{Pi0b}\\[0.3mm]
\Pi_2&=&\frac{1}{\cos\delta_2}\,\sum_i\,y_i(\mu)\,\mbox{Re}\langle 
Q_i\rangle_2\,.\label{Pi2b}
\end{eqnarray}
The factor $1/\hspace{-0.4mm}\cos\delta_I$ enhances the $\Delta I=1/2$ 
term in Eq.~(\ref{epspsm}) by about $25\,\%$ with respect to the $\Delta 
I=3/2$ one and consequently makes the cancellation between the $Q_6$ 
and $Q_8$ operators even less effective. It allows us to estimate the 
effect of multiple ($\pi-\pi$) rescattering on the imaginary part. In 
the second case, we assume zero phases and use the equations:
\begin{eqnarray}
\Pi_0&=&\,\sum_i\,y_i(\mu)\,\mbox{Re}\langle 
Q_i\rangle_0\,(1-\Omega_{\eta+\eta'})\,,\label{Pi0b2}\\[0.3mm]
\Pi_2&=&\,\sum_i\,y_i(\mu)\,\mbox{Re}\langle 
Q_i\rangle_2\,.\label{Pi2b2}
\end{eqnarray}
The comparison of the two cases provides, in part, an estimate for 
higher order effects. The latter case gives numerical results very close 
to those we would get if we used the imaginary parts from Tabs.~\ref{tab1} 
and~\ref{tab2}. As we already mentioned, we extract the values of 
$B_1^{(3/2)}$ and $B_2^{(3/2)}$ from the experimental value for
$\mbox{Re}A_2$. This procedure has also been followed in the 
phenomenological approach of the Munich group (last reference 
of~\cite{BJM}). Then
\begin{equation}
\mbox{Re}\langle Q_1\rangle_2\,(\mu)\,=\,\mbox{Re}\langle Q_2
\rangle_2\,(\mu)\,=\,\frac{\sqrt{2}\cos\delta_2}{G_F\,V_{ud}V_{us}}
\,\frac{\mbox{Re}A_2}{(z_1+z_2)\,(\mu)}\,=\,\frac{8.42\cdot 10^6\,
\mbox{MeV}^3}{(z_1+z_2)\,(\mu)}\,,\label{phq2}
\end{equation}
with $\mbox{Re}A_2=1.5\cdot 10^{-5}\,\mbox{MeV}$. The values of $z_1$ 
and $z_2$, for $600\,\mbox{MeV}\leq\mu\leq 1\,\mbox{GeV}$, are listed 
in Appendix~A of Ref.~\cite{hks}. All other $B$ factors are taken from 
Tabs.~\ref{tab3} and~\ref{tab4}. In particular, for $B_1^{(1/2)}$ 
and $B_2^{(1/2)}$ we use the values listed in Tab.~\ref{tab3}. These 
numbers were obtained in Ref.~\cite{hks} where it was shown that they
saturate the observed value of Re$A_0$ and are in good agreement with 
the phenomenological result of Ref.~\cite{BJM}.\footnote{Even though 
not fully consistent from a theoretical point of view, the values for 
$B_{1,2}^{(1/2)}$ in Tab.~\ref{tab3} can be used together with the 
experimental value for Re$A_0$ in the prefactor of Eq.~(\ref{epspsm}),
since the numbers in the table produce a value for Re$A_0$ close to the 
experimental one~\cite{hks}. In addition, the effect of $B_{1,2}^{(1/2)}$ 
in $\varepsilon'/\varepsilon$ is rather small.}
We note that $Q_1$ and $Q_2$ do not give a direct contribution to 
$\varepsilon'/\varepsilon$ since $y_1$ and $y_2$ are zero. Rather, 
$\langle Q_1\rangle_{0,2}$ and $\langle Q_2\rangle_{0,2}$ are used to 
sum up the contributions from the operators $Q_4$, $Q_9$, and $Q_{10}$ 
which are redundant below the charm threshold (see below).

The elimination of the scale dependence of QCD in the numerical result 
is an important criterion and we discuss it in some detail. The Wilson
coefficients in the effective hamiltonian in Eq.~(\ref{ham}) depend 
on the renormalization scale $\mu$. This should be matched with the 
scale dependence of the chiral operators and their respective matrix
elements. The bosonization of the density-density operators introduces 
masses which are also scale dependent. In particular, $\langle Q_6
\rangle_0$ and $\langle Q_8\rangle_2$ are proportional to $R^2\,=\,
\big[2m_K^2/(m_s+m_d)\big]^2\,\simeq\,4m_K^4/m_s^2$ which brings in 
a $\mu$ dependence through the quark masses already for the tree level 
(factorizable) contributions. This is different from the matrix elements 
of the current-current operators which are $\mu$ independent in the 
large-$N_c$ limit. In the products of $y_6$ and $y_8$ with the 
corresponding matrix elements, the $\mu$ dependence from the running 
quark mass is exactly cancelled by the diagonal evolution of the Wilson 
coefficients taken in the large-$N_c$ limit~\cite{BBG2,ab98}. This 
property is preserved at the two-loop level~\cite{BJM}. Furthermore, 
the $\mu$ dependence beyond the $m_s$ evolution, i.e., the $\mu$ 
dependence of $B_6^{(1/2)}$~and $B_8^{(3/2)}$ was shown in QCD to 
be only very weak for values above $1\,\mbox{GeV}$~\cite{BJM}. This 
requires that the (non-factorizable) $1/N_c$ corrections to the matrix 
elements of the $Q_6$ and $Q_8$ operators (which produce the scale 
dependence of the $B$ factors) should not show a large dependence on
the cutoff scale. The fact that the ${\cal O}(p^0/N_c)$ terms in 
Eqs.~(\ref{rm3}) and~(\ref{rm4}) have only a logarithmic cutoff 
dependence is for this reason welcome. Finally, the decrease of both 
$B$~factors with $\Lambda_c=\mu$ in Tabs.~\ref{tab3} and~\ref{tab4} which 
is due to these logarithms is qualitatively consistent with their $\mu$ 
dependence found for $\mu\geq\,1\,\mbox{GeV}$ in Ref.~\cite{BJM}, i.e., 
it has the correct slope. As shown below the residual scale dependence 
of $B_6^{(1/2)}$ and $B_8^{(3/2)}$ even if moderate is still too large to 
allow an exactly scale independent result for $\varepsilon'/\varepsilon$.

Throughout the numerical analysis of direct CP violation we take 
$\Lambda_{\mbox{\tiny QCD}}=\Lambda^{(4)}_{\overline{\mbox{\tiny MS}}}
=325\pm 80\,\mbox{MeV}$ corresponding roughly to $\alpha_s(M_Z)=0.118 
\pm 0.005$. For $\Omega_{\eta+\eta'}$ we adopt the range given after
Eq.~(\ref{Pi2}). The status of the strange quark mass has been 
reviewed recently in Refs.~\cite{ab98,bosch}, and we use the range             
\begin{equation}                                      
m_s\,(1\,\mbox{GeV})\,=\,150\,\pm\,25\,\,\mbox{MeV}\,,
\end{equation}                                        
which is in the ball park of the values obtained in the quenched 
lattice calculations (see Ref.~\cite{gupta} and references therein; 
for a very recent analysis see Ref.~\cite{ghsw}) and from QCD sum rules 
\cite{qcdsum}. We note that the QCD sum rule results are generally higher 
than the lattice values. Lower bounds on the strange mass have been derived 
in Ref.~\cite{msbound}.
\renewcommand{\topfraction}{1.0}
\renewcommand{\textfraction}{0.0}
\noindent
\begin{figure}[tbh]
\vspace*{-1cm}
\centerline{\epsfig{file=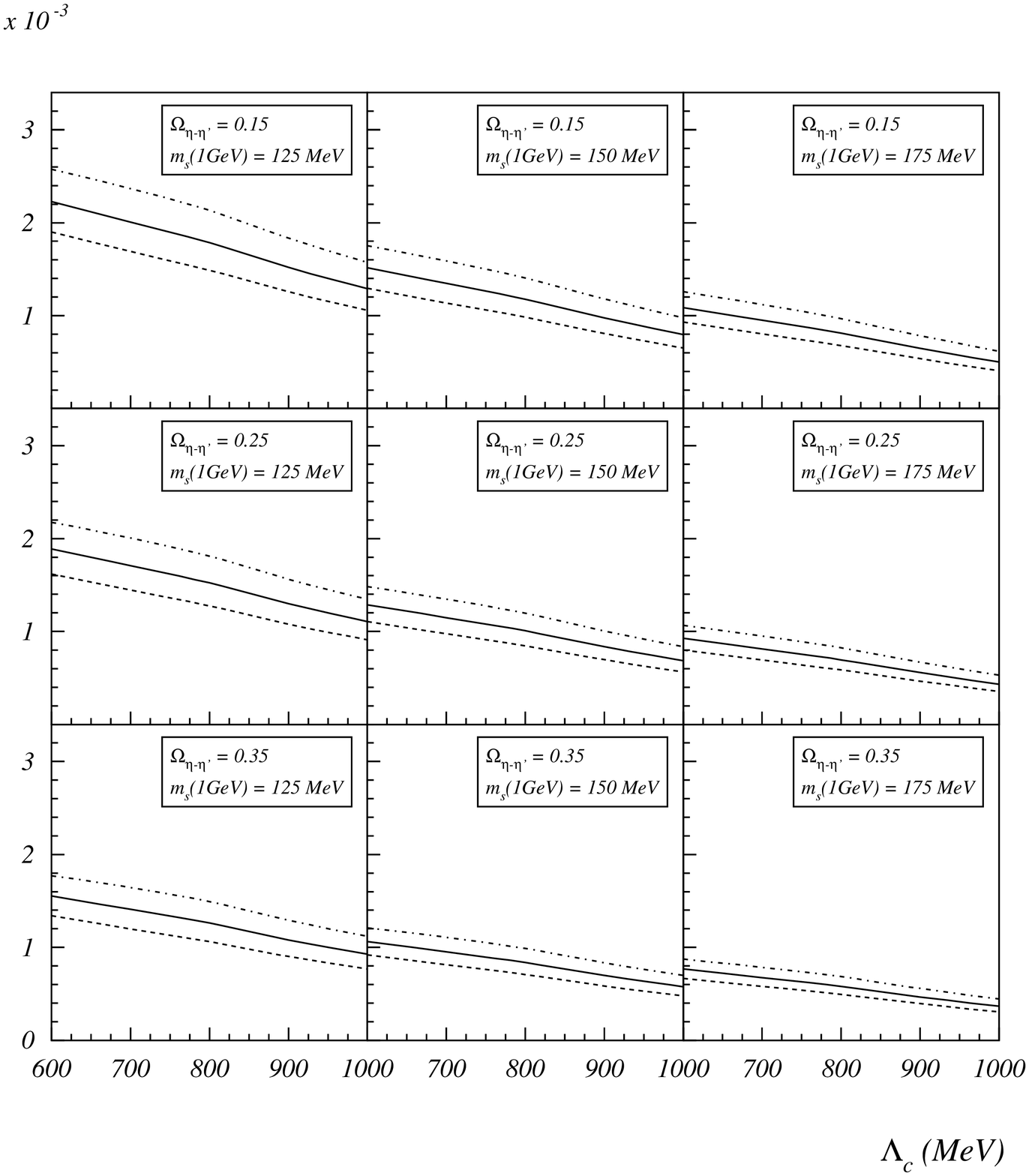,width=17.2cm}}
\caption{$\varepsilon'/\varepsilon$ using LO Wilson coefficients and the
experimental phases, plotted for various values of $m_s(1\,\mbox{GeV})$ 
and $\Omega_{\eta+\eta'}$ as a function of the matching scale $\lc=\mu$.
We use $\mbox{Im}\lambda_t=1.33\cdot 10^{-4}$. The solid (dashed, 
dot-dashed) lines correspond to $\Lambda_{\mbox{\tiny QCD}}=325$ 
$(245,\,405)\,\,\mbox{MeV}$.\label{epsp1}}
\end{figure}
\renewcommand{\topfraction}{1.0}
\renewcommand{\textfraction}{0.0}
\noindent
\begin{figure}[t]
\centerline{\epsfig{file=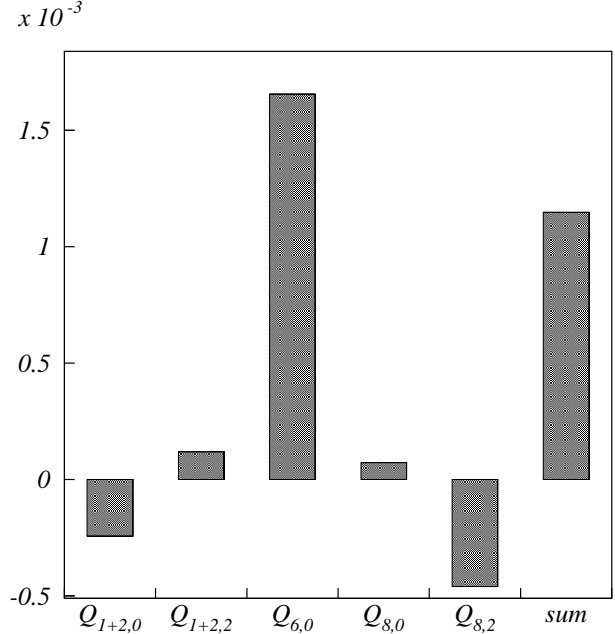,height=9.1cm}}
\vspace*{2mm}
\caption{The $I=0$ and $I=2$ contributions to $\varepsilon'/\varepsilon$ 
of the dominant operators using LO Wilson coefficients and the experimental 
phases. We also use the central values for $\Lambda_{\mbox{\tiny QCD}}$, 
$m_s(1\,\mbox{GeV})$, $\Omega_{\eta+\eta'}$, and $\mbox{Im}\lambda_t$ at 
a scale of $\Lambda_c=700\,$MeV. The contributions of the operators $Q_3$ 
and $Q_5$ are negligible and are not included in the figure.\label{sky1}}
\end{figure}
\renewcommand{\topfraction}{1.0}
\renewcommand{\textfraction}{0.0}
\noindent
\begin{figure}[t]
\centerline{\epsfig{file=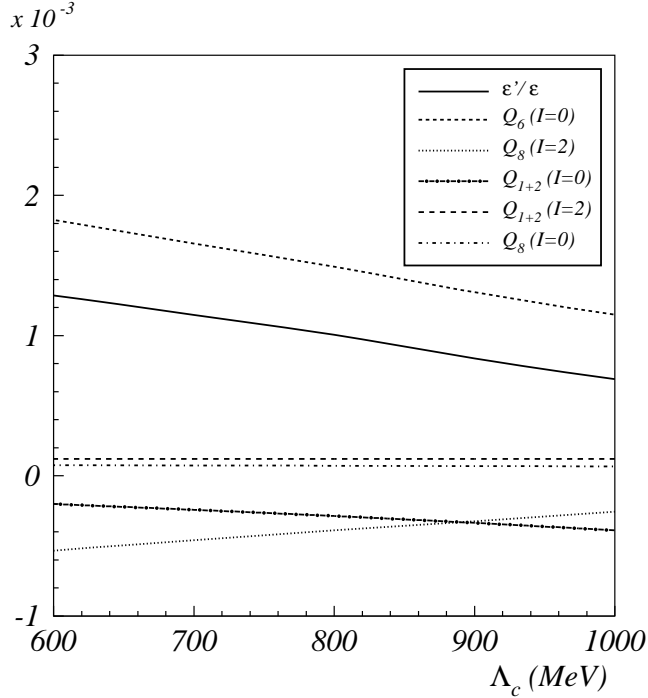,height=9.5cm}}
\vspace*{3mm}
\caption{Various contributions to $\varepsilon'/\varepsilon$ using LO 
Wilson coefficients and the experimental phases, plotted as functions 
of the matching scale $\Lambda_c$. We use central values for $m_s$, 
$\Omega_{\eta+\eta'}$, $\Lambda_{\mbox{\tiny QCD}}$, and $\mbox{Im}
\lambda_t$.\label{op1}}
\end{figure}
\renewcommand{\topfraction}{1.0}
\renewcommand{\textfraction}{0.0}
\noindent
\begin{figure}[t]
\vspace*{-1cm}
\centerline{\epsfig{file=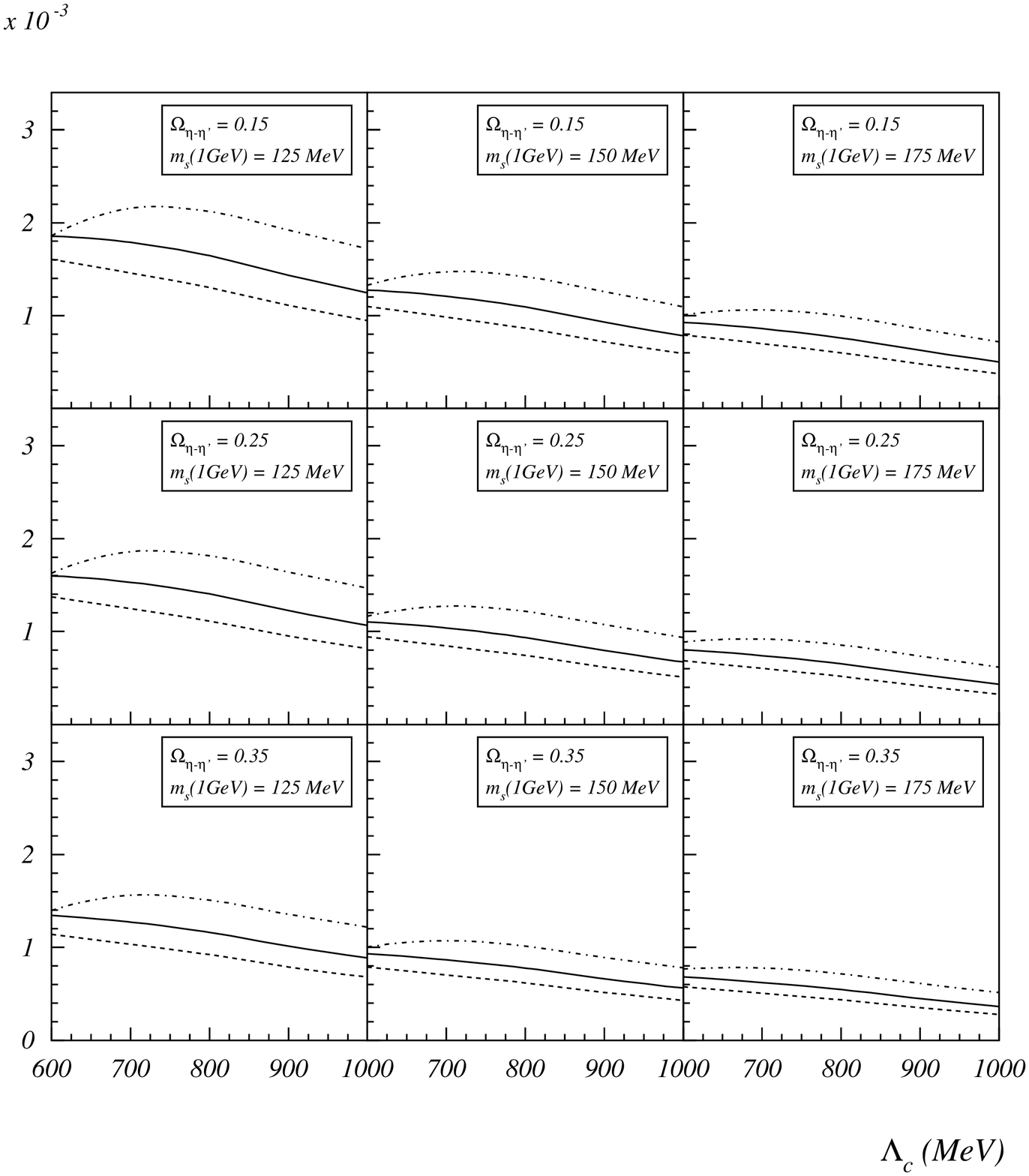,width=17.2cm}}
\caption{Same as in Fig.~\ref{epsp1}, now using NLO Wilson coefficients
in the NDR scheme. The solid (dashed, dot-dashed) lines correspond to
$\Lambda_{\mbox{\tiny QCD}}=\Lambda^{(4)}_{\overline{\mbox{\tiny
MS}}}=325$ $(245,\,405)\,\,\mbox{MeV}$.\label{epsp2}}
\end{figure}
\renewcommand{\topfraction}{1.0}
\renewcommand{\textfraction}{0.0}
\noindent
\begin{figure}[t]
\centerline{\epsfig{file=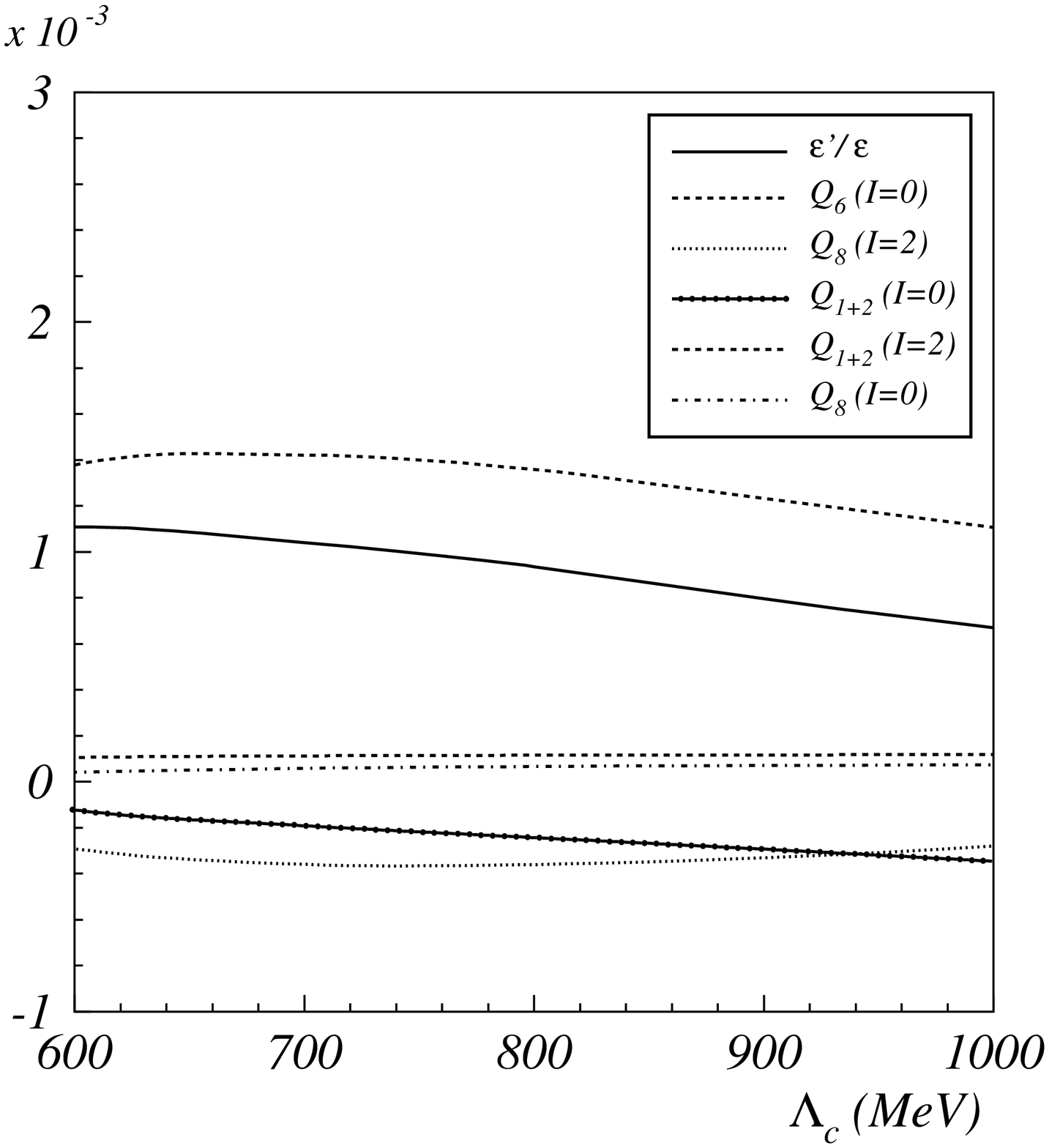,width=9.5cm}}
\vspace*{1mm}
\caption{Same as in Fig.~\ref{op1}, now using NLO Wilson coefficients
in the NDR scheme.\label{op2}}
\end{figure}
\renewcommand{\topfraction}{1.0}
\renewcommand{\textfraction}{0.0}
\noindent
\begin{figure}[t]
\centerline{\epsfig{file=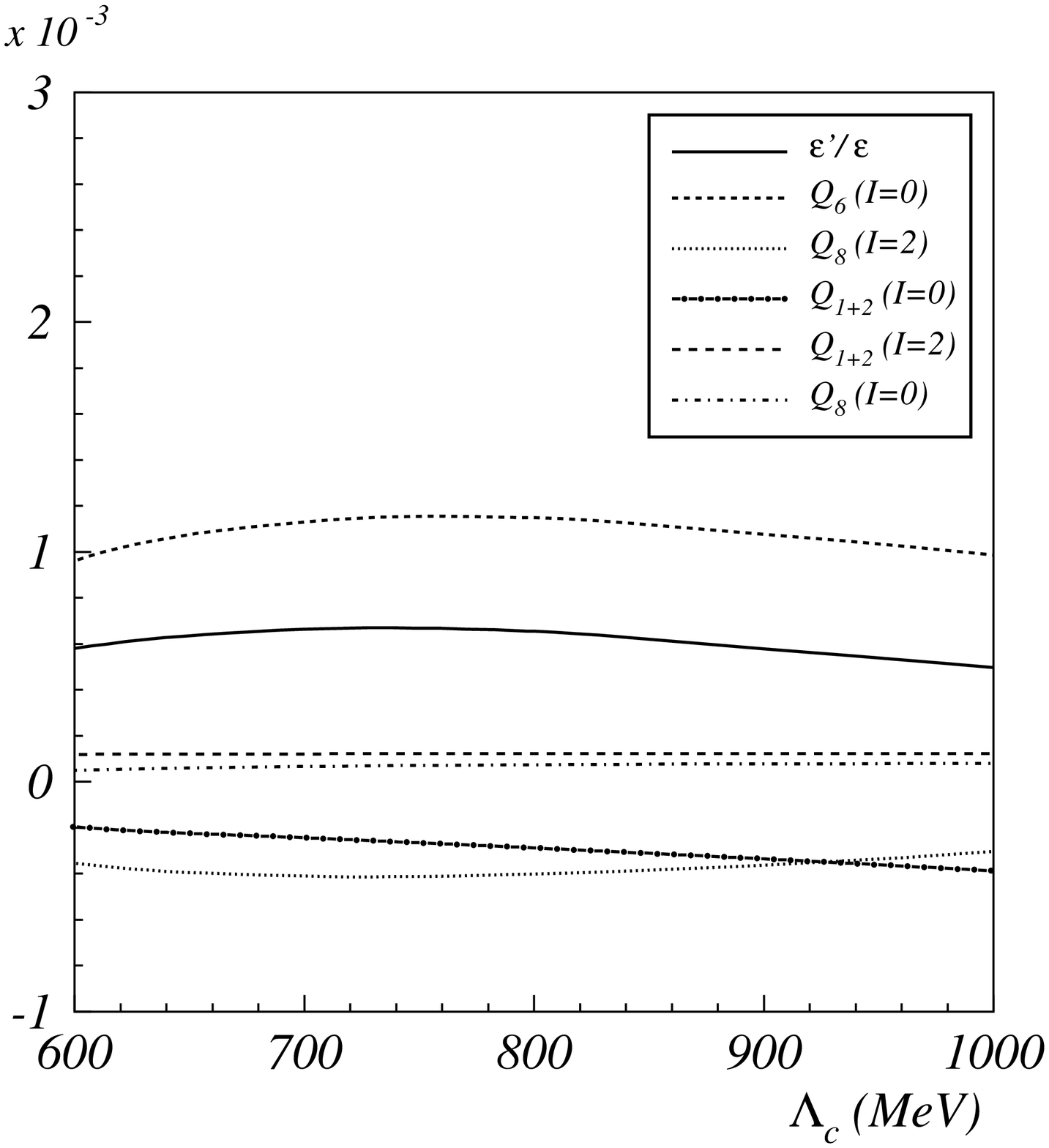,width=9.5cm}}
\vspace*{1mm}
\caption{
Same as in Fig.~\ref{op1}, now using NLO Wilson coefficients in the HV 
scheme.\label{op3}}
\end{figure}
\renewcommand{\topfraction}{1.0}
\renewcommand{\textfraction}{0.0}
\noindent
\renewcommand{\topfraction}{1.0}
\begin{figure}[t]
\vspace*{-1cm}
\centerline{\epsfig{file=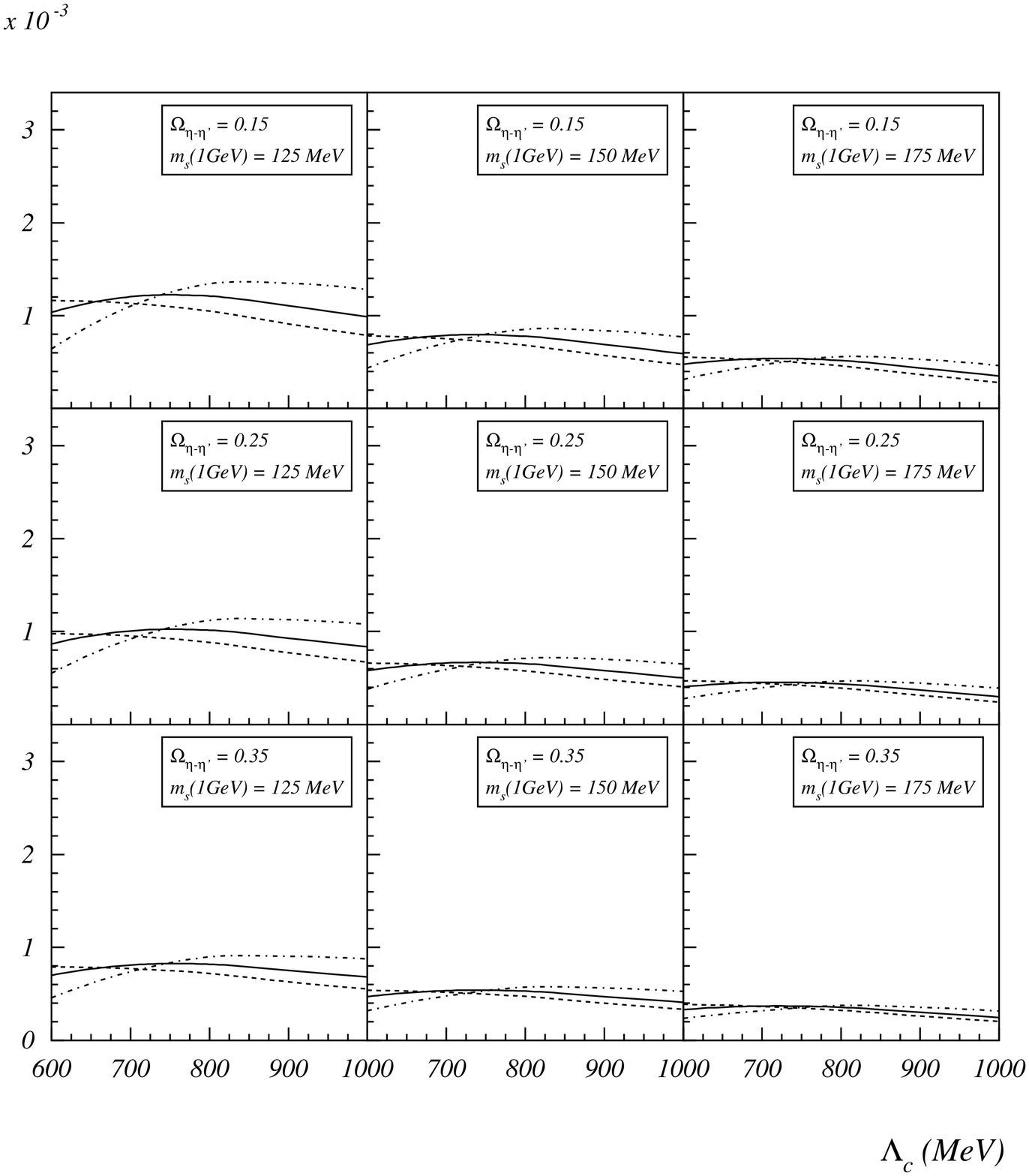,width=17.2cm}}
\caption{Same as in Fig.~\ref{epsp1}, now using NLO Wilson coefficients 
in the HV scheme. The solid (dashed, dot-dashed) lines correspond to
$\Lambda_{\mbox{\tiny QCD}}=\Lambda^{(4)}_{\overline{\mbox{\tiny
MS}}}=325$ $(245,\,405)\,\,\mbox{MeV}$.\label{epsp3}}
\end{figure}

In Fig.~\ref{epsp1} we depict $\varepsilon'/\varepsilon$ as a function of 
the matching scale ($\mu=\Lambda_c$), calculated from Eqs.~(\ref{Pi0b}) 
and~(\ref{Pi2b}) with LO Wilson coefficients for the central value of 
$\mbox{Im}\lambda_t$ and for various values of $m_s$, $\Omega_{\eta+\eta'}$, 
and $\Lambda_{\mbox{\tiny QCD}}$ according to their ranges defined above. 
For low values of the matching scale we find a rather moderate enhancement 
of the VSA result which is due to the weaker cancellation between the $Q_6$ 
and $Q_8$ operators. However, one might note that very large values for 
$\varepsilon'/\varepsilon$ in the range of the recent Fermilab measurement 
\cite{ktev} are not reached with the $B$ factors listed in Tabs.~\ref{tab3} 
and~\ref{tab4} together with Eq.~(\ref{phq2}), if central values are used 
for the parameters. Indeed, adopting central values for $m_s$, 
$\Omega_{\eta+\eta'}$, $\Lambda_{\mbox{\tiny QCD}}$, and $\mbox{Im}
\lambda_t$ and varying $\Lambda_c$ between 600 MeV and 900 MeV we 
obtain as `central range' for the CP ratio:
\begin{equation}
8.4\cdot 10^{-4}\,\,\leq\,\,\varepsilon'/\varepsilon\,\,(\mbox{LO-central})
\,\,\leq\,\,12.9\cdot 10^{-4}\,.\label{epscentreslo}
\end{equation}
This is also illustrated in Fig.~\ref{sky1} where we show the various 
contributions to $\varepsilon'/\varepsilon$ for central values of the
parameters at a scale of $\Lambda_c=700\,$MeV. For this value of the 
cutoff $B_6^{(1/2)}$ is very close to unity whereas $B_8^{(3/2)}$ 
is significantly suppressed which leads to a value for $\varepsilon'/
\varepsilon\,$ of $\,11.5\,\cdot 10^{-4}$. Smaller numbers are obtained 
for larger values of the cutoff. Another noticeable contribution, beside 
that of $Q_8$, which reduces the value of $\varepsilon'/\varepsilon$ is 
the $I=0$ component of $Q_1$ and $Q_2$. As we already mentioned above, 
this contribution comes from the $Q_4$, $Q_9$, and $Q_{10}$ operators which 
are redundant below the charm threshold and satisfy, to LO and at NLO in 
the HV scheme, the relations in Eq.~(\ref{linear-op}). In the NDR scheme 
the relations receive small ${\cal O}(\alpha_s)$ and ${\cal O}(\alpha)$ 
corrections~\cite{BJM}.

In Fig.~\ref{op1} we show how the various terms depend on the choice 
of the matching scale. In particular, we observe that the behaviour of 
$\varepsilon'/\varepsilon$ is almost identical to the one of $y_6\,\langle 
Q_6\rangle_0$. This is due to the fact that the ratio $B_6^{(1/2)}/B_8^{(3/2)}$ 
is approximately stable over the whole range of the cutoff $\Lambda_c$ 
and, consequently, $y_6\,\langle Q_6\rangle_0$ and $(-)\,y_8\,\langle 
Q_8\rangle_2$ fall off roughly in the same way. We note that the ratio 
$y_6(\mu)/m_s^2(\mu)$ increases by about $12\,\%$ if the scale 
$\mu=\Lambda_c$ is varied between $600\,$MeV and $1\,$GeV, whereas 
$B_6^{(1/2)}(\Lambda_c)$ decreases by $44\,\%$. A similar statement 
applies to $Q_8$. The decrease of $B_6^{(1/2)}$ and $B_8^{(3/2)}$ is 
therefore qualitatively consistent with the (non-diagonal) evolution of 
$y_6$ and $y_8$ computed in the leading logarithmic approximation, and it 
leads to a fairly moderate overall scale dependence. This property is due 
to the fact that the ${\cal O}(p^0/N_c)$ terms in Eqs.~(\ref{rm3}) and 
(\ref{rm4}) have only a logarithmic cutoff dependence which, nevertheless, 
still goes beyond the $\mu$ dependence of the short-distance part. In 
this situation it would be tempting to adopt the large-$N_c$ values for 
$B_6^{(1/2)}$ and $B_8^{(3/2)}$ which are scale independent and coincide, 
to a very good approximation, with their VSA values $B_6^{(1/2)}=B_8^{(3/2)}
=1$. However, the results show that $1/N_c$ corrections are important, and 
to recover the VSA values would require an a priori unexpected cancellation 
of the ${\cal O}(p^0/N_c)$ corrections with higher order terms or 
contributions from higher resonances. Therefore, the VSA might 
underestimate the true range of uncertainty in the analysis of 
$\varepsilon'/\varepsilon$.

The dependences of the result on $m_s$, $\Omega_{\eta+\eta'}$, and 
$\Lambda_{\mbox{\tiny QCD}}$ are given in Fig.~\ref{epsp1}. Among them 
the $m_s$ dependence is the most important one. The dependence on 
Im$\lambda_t$, to a large degree of accuracy \cite{bosch}, is multiplicative 
and can be obtained in straightforward way. If we take into account the 
residual dependence on the matching scale by varying $\mu=\Lambda_c$ 
between $600$ and $900\,$MeV and scan independently the theoretical input 
parameters and the experimentally measured numbers, we obtain the following 
range for $\varepsilon'/\varepsilon$ calculated with LO Wilson coefficients:
\begin{equation}
3.1\cdot 10^{-4}\,\,\leq\,\,\varepsilon'/\varepsilon\,\,(\mbox{LO})
\,\,\leq\,\,31.6\cdot 10^{-4}\,.\label{epsreslo}
\end{equation}
The quoted range results from a variation of $m_s$, $\Omega_{\eta+\eta'}$, 
and $\Lambda_{\mbox{\tiny QCD}}$ in Eqs.~(\ref{Pi0b}) and~(\ref{Pi2b})
as depicted in Fig.~\ref{epsp1} and also allows for a variation of 
$\mbox{Im}\lambda_t$ according to the range defined in~Eq.~(\ref{ltval}).

We investigate next the dependence on the NLO Wilson coefficients. The 
NLO values are scheme dependent and are calculated within naive dimensional 
regularization (NDR) and in the 't Hooft-Veltman scheme (HV), respectively.
As already mentioned, the absence of any reference to the renormalization 
scheme dependence in the low-energy calculation prevents a complete matching 
at the next-to-leading order~\cite{ab98}. Nevertheless, a comparison of the 
numerical results obtained from the LO and NLO coefficients is useful in 
order to estimate the corresponding uncertainties and to test the validity 
of perturbation theory. 
\begin{table}[t]
\begin{eqnarray*}
\begin{array}{|c||c|c|}\hline
 & \mbox{Case}\,\, 1& \mbox{Case}\,\, 2\\
\hline\hline 
\rule{0cm}{5mm} 
\mbox{LO}& \,\,8.4 \,\,\leq\,\,\varepsilon'/\varepsilon\,\, 
\leq\,\,12.9 \,\, 
& \,\,6.3 \,\,\leq\,\,\varepsilon'/\varepsilon\,\, 
\leq\,\,9.5 \,\, \\[0.5mm]
\mbox{NDR}& \,\,8.0 \,\,\leq\,\,\varepsilon'/\varepsilon\,\, 
\leq\,\,11.0 \,\, 
& \,\,5.9 \,\,\leq\,\,\varepsilon'/\varepsilon\,\,
\leq\,\,8.4 \,\, \\[0.5mm]
\mbox{HV}& \,\,5.8 \,\,\leq\,\,\varepsilon'/\varepsilon\,\, 
\leq\,\,6.6\,\,\,\, 
& \,\,4.2 \,\,\leq\,\,\varepsilon'/\varepsilon\,\,
\leq\,\,4.7 \,\, \\[0.5mm]
\hline
\,\mbox{LO}\,+\,\mbox{NDR}\,+\,\mbox{HV}\,
& \,\,5.8 \,\,\leq\,\,\varepsilon'/\varepsilon\,\,\leq \,\,12.9\,\, 
& \,\,4.2 \,\,\leq\,\,\varepsilon'/\varepsilon\,\,
\leq\,\,9.5\,\,\\[0.5mm]
\hline
\end{array}
\end{eqnarray*}
\caption{Central ranges for $\varepsilon'/\varepsilon$ (in units of
$10^{-4}$) at LO and NLO (NDR and HV). The numbers are obtained for
central values of $m_s$, $\Omega_{\eta+\eta'}$, Im$\lambda_t$, and
$\Lambda_{\mbox{\tiny QCD}}$ by varying $\Lambda_c$ between 600 and 
$900\,\mbox{MeV}$. `Case 1' and `Case 2' correspond to the use of 
Eqs.~(\ref{Pi0b})\,-\,(\ref{Pi2b}) and (\ref{Pi0b2})\,-\,(\ref{Pi2b2}),
respectively.\label{tab5}}
\end{table}
\begin{table}[t]
\begin{eqnarray*}
\begin{array}{|c||c|c|}\hline
 & \mbox{Case}\,\, 1& \mbox{Case}\,\, 2\\
\hline\hline 
\rule{0cm}{5mm} 
\mbox{LO}& \,\,3.1 \,\,\leq\,\,\varepsilon'/\varepsilon\,\, 
\leq\,\,31.6 \,\, 
& \,\, 2.4\,\,\leq\,\,\varepsilon'/\varepsilon\,\, 
\leq\,\,23.2 \,\, \\[0.5mm]
\mbox{NDR}& \,\,2.7 \,\,\leq\,\,\varepsilon'/\varepsilon\,\, 
\leq\,\,26.4 \,\, 
& \,\,2.1 \,\,\leq\,\,\varepsilon'/\varepsilon\,\,
\leq\,\,20.2 \,\, \\[0.5mm]
\mbox{HV}& \,\,1.9 \,\,\leq\,\,\varepsilon'/\varepsilon\,\, 
\leq\,\,16.5 \,\, 
& \,\,1.5 \,\,\leq\,\,\varepsilon'/\varepsilon\,\,
\leq\,\,11.9 \,\, \\[0.5mm]
\hline
\,\mbox{LO}\,+\,\mbox{NDR}\,+\,\mbox{HV}\,
& \,\,1.9 \,\,\leq\,\,\varepsilon'/\varepsilon\,\,\leq \,\,31.6\,\, 
& \,\,1.5 \,\,\leq\,\,\varepsilon'/\varepsilon\,\,
\leq\,\,23.2\,\,\\[0.5mm]
\hline
\end{array}
\end{eqnarray*}
\caption{Same results as in Tab.~\ref{tab5} but for the complete 
scanning of the parameters ($\Lambda_c$, $m_s$, $\Omega_{\eta+\eta'}$ and 
$\Lambda_{\mbox{\tiny QCD}}$, and Im$\lambda_t$) as explained in the text. 
\label{tab6}}
\end{table}

In the NDR scheme, introducing the NLO coefficients does not 
noticeably affect our numerical results (see Fig.~\ref{epsp2}). For 
$\Lambda^{(4)}_{\overline{\mbox{\tiny MS}}}=\Lambda_{\mbox{\tiny QCD}}
\lesssim 325\,\mbox{MeV}$ we find slightly lower values for $\varepsilon'/
\varepsilon$ and a somewhat larger difference between the results obtained 
for $\Lambda_{\mbox{\tiny QCD}}=325\,\mbox{MeV}$ and $405\,\mbox{MeV}$, 
respectively. Generally, the difference between LO and NLO is more 
pronounced for very low values of the matching scale, but it is still 
moderate except for $\Lambda_{\mbox{\tiny QCD}}=405\,\mbox{MeV}$. For 
$\Lambda_{\mbox{\tiny QCD}}=325\,\mbox{MeV}$ ($245\,\mbox{MeV}$) the effect 
of the NLO coefficients is rather small, and values for the matching scale 
as low as $600\,$-$\,700\,\mbox{MeV}$ appear to be acceptable. We also 
notice a slightly smaller scale dependence, that is to say, the NLO Wilson 
coefficients further improve the stability of the calculation. This property 
becomes obvious if we investigate the various contributions to $\varepsilon'/
\varepsilon$ (compare Figs.~\ref{op1} and~\ref{op2}). In particular, at NLO 
in the NDR scheme we observe a smaller variation of $y_6\,\langle Q_6
\rangle_0$ and $y_8\,\langle Q_8\rangle_2$ in the range of $\Lambda_c$ 
between $600\,\mbox{MeV}$ and $1\,\mbox{GeV}$. Nevertheless, the numerical 
effect of the NLO coefficients is rather moderate, and the `central' and 
scanned ranges quoted in Tabs.~\ref{tab5} and~\ref{tab6} are close to the 
LO results given in Eqs.~(\ref{epscentreslo}) and~(\ref{epsreslo}).

In the HV scheme, the effect of the NLO coefficients is more pronounced. 
Both $y_6\,\langle Q_6\rangle_0$ and $y_8\,\langle Q_8\rangle_2$ are rather 
stable over a large range of the matching scale leading to an approximately 
stable result for $\varepsilon'/\varepsilon$ between $700\,\mbox{MeV}$ and 
$1\,\mbox{GeV}$. This is shown in Fig.~\ref{op3} for the central values of 
$m_s(1\,\mbox{GeV})$, $\Omega_{\eta+\eta'}$, $\mbox{Im}\lambda_t$, and
$\Lambda_{\mbox{\tiny QCD}}$. On the other hand, for $\Lambda_{\mbox{\tiny 
QCD}}=405\,\mbox{MeV}$ and $\Lambda_c\lesssim 700\,\mbox{MeV}$ we observe a
noticeable slope indicating the breakdown of the perturbative expansion of 
QCD (see Fig.~\ref{epsp3}). However, for moderate values of the matching 
scale the numerical values for the ratio depend weakly on the choice of 
$\Lambda_{\mbox{\tiny QCD}}$ (see Fig.~\ref{epsp3}), which makes the result 
rather stable. Generally, at NLO in the HV scheme we obtain smaller values 
for $\varepsilon'/\varepsilon$ (see Tabs.~\ref{tab5} and~\ref{tab6}).

We note that at NLO the maximum value for the ratio is found for moderate 
values of $\Lambda_c$ around $700\,$-$\,800\,\mbox{MeV}$, whereas the upper 
bound in Eq.~(\ref{epsreslo}) refers to low values of the matching scale 
($\sim 600\,\mbox{MeV}$). Finally, one might note that the numerical values 
of the Wilson coefficients in the HV scheme communicated to us by M.~Jamin
\cite{jam} correspond to the treatment of the two-loop anomalous dimensions 
used in Ref.~\cite{BJM} which differs from the one used in Ref.~\cite{CFMR}. 
For this reason the NLO corrections to the Wilson coefficients in the HV 
scheme presented in Appendix~\ref{append} are generally smaller than the 
ones found in~Ref.~\cite{CFMR}~(for a discussion of this point see also 
Ref.~\cite{ab98}). 

So far in the numerical analysis we have used 
Eqs.~(\ref{Pi0b})\,-\,(\ref{Pi2b}) together with the 
phenomenological values for the phases~\cite{phases}. Replacing 
them by Eqs.~(\ref{Pi0b2})\,-\,(\ref{Pi2b2}) leads to lower values 
for $\varepsilon'/\varepsilon$ (see Case\,\,2 in Tabs.~\ref{tab5}
and~\ref{tab6}). The numerical results are very close to those we would 
get if we used the imaginary parts obtained at the one-loop level in the 
$1/N_c$ approach. Both the central values and the upper bounds in the 
scanned ranges for $\varepsilon'/\varepsilon$ are lower due to a smaller 
contribution from the $\Delta I =1/2$ terms. However, the modifications 
do not change substantially our picture of $\varepsilon'/\varepsilon$. 
As mentioned above, the comparison of the two cases provides, in part, 
an estimate for higher order effects. 

In conclusion, the fact that we use rather low values for the matching 
scale makes some of the Wilson coefficients rather sensitive to NLO
corrections. In particular, $y_6$ and $y_8$ depend noticeably on the 
choice of the $\gamma_5$ scheme in dimensional regularization. For 
example, for $\mu=700\,\mbox{MeV}$ and $\Lambda_{\mbox{\tiny QCD}}
=325\,$MeV the values of $y_6$ and $y_8$ at LO and in the NDR and HV 
schemes differ approximately by 20\,-\,30\,\%. Since the non-perturbative 
calculation of the matrix elements is insensitive to this dependence, 
the corresponding uncertainty must be included in the final result for 
$\varepsilon'/\varepsilon$. Collecting together the LO and NLO values 
in Tab.~\ref{tab5} from Eqs.~(\ref{Pi0b})\,-\,(\ref{Pi2b}) and
(\ref{Pi0b2})\,-\,(\ref{Pi2b2}) we get the following range:
\begin{equation}  
4.2\cdot 10^{-4}\,\,\leq\,\,\varepsilon'/\varepsilon\,\,(\mbox{central})
\,\,\leq\,\,12.9\cdot 10^{-4}\,,\label{epscenttot}
\end{equation}
which, for central values of $m_s$, $\Omega_{\eta+\eta'}$, 
$\Lambda_{\mbox{\tiny QCD}}$, and Im$\lambda_t$, takes into account the 
theoretical errors inherent to the method (dependence on the scheme and 
matching scale). Furthermore, it includes the expected errors due to the 
neglect of higher order corrections to the imaginary part. Similarly 
collecting the values in Tab.~\ref{tab6} we obtain the following range 
from the complete scanning of the parameters:
\begin{equation}  
1.5\cdot 10^{-4}\,\,\leq\,\,\varepsilon'/\varepsilon
\,\,\leq\,\,31.6\cdot 10^{-4}\,,\label{epstot}
\end{equation}
which also takes into account the uncertainties in the values 
for $m_s$, $\Omega_{\eta+\eta'}$, $\Lambda_{\mbox{\tiny QCD}}$, and 
Im$\lambda_t$.\footnote{A comparison with the results of other calculations 
performed within a specific scheme and treatment of the final state interaction 
phases should be done using the numbers in Tab.~\ref{tab6}.} The upper bound 
from our calculation in Eq.~(\ref{epstot}) is rather close to the central 
value of the new Fermilab measurement~\cite{ktev} and requires a conspiracy 
of the parameters within their ranges of uncertainties given above.

The present world average for the ratio including earlier 
measurements is $\mbox{Re}(\varepsilon'/\varepsilon)=(21.8\pm 3.0)\cdot 
10^{-4}$~\cite{ktev}. Our result indicates that the experimental data 
can be accommodated in the standard model. A major uncertainty in the 
theoretical estimate of $\varepsilon'/\varepsilon$ is due to the choice 
of $m_s$, which enters the calculation through the matrix elements of 
the operators $Q_6$ and $Q_8$ [see Eqs.~(\ref{rm3}) and~(\ref{rm4})]. In 
Eq.~(\ref{epstot}) we have taken $m_s(1\,\mbox{GeV})=150\pm25\,\mbox{MeV}$ 
which is in the range of the values obtained in quenched lattice calculations 
and from the QCD sum rules. Adopting even lower values for $m_s$ would allow 
us to relax the upper bound quoted above. However, recently the ALEPH 
collaboration analyzed the measured mass spectra of the strange $\tau$ 
decay modes and reported a value of $m_s(1\,\mbox{GeV})=234{+61\atop -76}
\,\mbox{MeV}$~\cite{aleph}. It will be interesting to see whether this 
large (central) value for $m_s$ will remain when the error is 
reduced. Very recently, the value $m_s(1\,\mbox{GeV})=(188\pm 22)\,$MeV
was obtained using a $\tau$-like decay sum rule for the $\phi$ meson 
\cite{narison}, which is consistent with the range used in this paper. 
The determination of $\mbox{Im}\lambda_t$ will be further improved by 
precision tests of the unitarity triangle \cite{ab98} removing to a large 
extent the corresponding uncertainty in the analysis of direct CP violation. 
$\Omega_{\eta+\eta'}$~which measures the contribution to $\varepsilon'/
\varepsilon$ from the isospin breaking in the quark masses was estimated 
in Ref.~\cite{BG1} in the large-$N_c$ limit, and it will be a challenge 
to investigate, in future studies, the $1/N_c$ corrections to this 
parameter. Finally, the calculation of the hadronic matrix elements 
even though largely improved by including $1/N_c$ corrections may still
be plagued by noticeable uncertainties. Our analysis so far included 
terms of ${\cal O}(p^0)$, ${\cal O}(p^0/N_c)$, and ${\cal O}(p^2)$ for 
the matrix elements of the density-density operators and terms of 
${\cal O}(p^2)$, ${\cal O}(p^2/N_c)$, and ${\cal O}(p^4)$ for the matrix 
elements of the current-current operators. In the following section we 
shall investigate the effect of higher order corrections. In particular 
we will consider the terms of ${\cal O}(p^2/N_c)$ for the matrix elements 
of $Q_6$.

%
\section{Higher Order Corrections \label{hoc}}
In the previous section we have shown that the calculation of the hadronic
matrix elements in the $1/N_c$ expansion leads to a well defined range of 
values for $\varepsilon'/\varepsilon$ which can account, to a large extent, 
for the weighted average of the experimental measurements 
\cite{barr,gibb,ktev}. However, the central values obtained are lower
than the values of the new measurement~\cite{ktev}. The upper bound from 
our calculation requires, within the standard model, specific values of 
various parameters. In particular, lower values of the strange quark mass 
are favoured. In our analysis so far we varied the theoretical input 
parameters independent of each other and considered the experimental 
results within one standard deviation. This conservative attitude may to 
some extent exaggerate the differences~\cite{BJL96}. In the present section 
we investigate the higher order corrections and consider in particular the 
question whether these corrections are able to substantially enhance the 
prediction for $\varepsilon'/\varepsilon$, so that a large value for the
ratio could be explained even for central values of the input parameters.

In the twofold expansion, the higher order corrections to the matrix 
elements of $Q_6$ and $Q_8$ are of orders: ${\cal O}(p^4)$, ${\cal O}
(p^0/N_c^2)$, and ${\cal O}(p^2/N_c)$. In this section we will consider 
the ${\cal O}(p^2/N_c)$ contribution which brings in, for the first time, 
quadratic corrections on the cutoff. From general counting arguments we 
show that these corrections are expected to be large for $Q_6$, which is 
a peculiar operator. $Q_6$ is consequently not protected from possible 
large corrections beyond the large-$N_c$ limit, and we cannot exclude 
the possibility that the contribution of $Q_6$ brings $\varepsilon'/
\varepsilon$ close to the experimental value for central values of the 
parameters. Calculating the ${\cal O}(p^2/N_c)$ correction for $Q_6$ in 
the chiral limit we explicitly find that it is indeed large and positive.

Before investigating the ${\cal O}(p^2/N_c)$ corrections we briefly 
estimate part of the higher order corrections replacing the `$1/F_\pi$ 
expansion' by a `$1/F_K$ expansion'. As already discussed in 
Ref.~\cite{HKPSB}, we could have used the ratio $1/f$ in place of $1/F_\pi$ 
in the next-to-leading order terms of Eqs.~(\ref{rm1})\,-\,(\ref{rm4}). This 
choice would be consistent at the level of first order corrections in the 
twofold series expansion, as the difference concerns higher order effects. 
However, the scale dependence of $f$ (which is mainly quadratic) is absorbed 
through the factorizable loops to the matrix elements at the next order in 
the parameter expansion and does not occur in the matching with the
short-distance contribution~\cite{HKPSB}. Consequently, it is more 
appropriate to choose the physical decay constant in the expressions 
under consideration. In this situation, we can use, instead of $F_\pi$, the 
kaon decay constant $F_K$ which gives an indirect estimate of higher order 
corrections. In Tab.~\ref{tab7} we show the effect of this modification, 
to ${\cal O}(p^0/N_c)$, on the values of $B_6^{(1/2)}$ and $B_8^{(3/2)}$. 
We notice that the numbers are generally larger for the `$1/F_K$ expansion'. 
In particular, for $\Lambda_c\lesssim 900\,\mbox{MeV}$ the $B_6^{(1/2)}$ 
factor is enhanced compared to the VSA value. This change, in spite of the 
somewhat smaller reduction of $B_8^{(3/2)}$, leads to a moderate enhancement 
of $\varepsilon'/\varepsilon$ which further improves the agreement with the 
observed value. Numerically, collecting together the various terms we
get $5.3\cdot 10^{-4}\,\leq\,\varepsilon'/\varepsilon\,(\mbox{central})
\,\leq\,15.8\cdot 10^{-4}$. Scanning independently the input parameters 
we obtain $1.8\cdot 10^{-4}\,\leq\,\varepsilon'/\varepsilon\,\leq\,
38.3\cdot 10^{-4}$ in place of Eq.~(\ref{epstot}). Adopting zero phases 
reduces the upper bound to $28.3\cdot 10^{-4}$.  
\begin{table}[t]
\begin{eqnarray*}
\begin{array}{|c||c|c|c|c|c|c|}\hline
\lc&0.6\,\,\mbox{GeV}&0.7\,\,\mbox{GeV}&
0.8\,\,\mbox{GeV}&0.9\,\,\mbox{GeV}&1.0\,\,\mbox{GeV} \\
\hline\hline
\rule{0cm}{5mm}
B_{6}^{(1/2)}&  1.10  &  0.96  &  0.84  &  0.72  &  0.62  \\
             & (1.30) & (1.19) & (1.09) & (0.99) & (0.91) \\[0.5mm]
B_{8}^{(3/2)}&  0.64  &  0.56  &  0.49  &  0.42  &  0.34  \\
             & (0.72) & (0.66) & (0.59) & (0.53) & (0.46) \\[0.5mm]
\hline
\end{array}
\end{eqnarray*}
\caption{Bag parameters $B_{6}^{(1/2)}$ and  $B_{8}^{(3/2)}$ shown for 
various values of $\Lambda_c$. The numbers in the parentheses are obtained
by replacing $F_\pi$ by $F_K$ in the next-to-leading order expression.
\label{tab7}}
\vspace*{-4mm}
\end{table}
\renewcommand{\topfraction}{1.0}
\renewcommand{\textfraction}{0.0}
\noindent
\begin{figure}[t]
\vspace*{-1cm}
\centerline{\epsfig{file=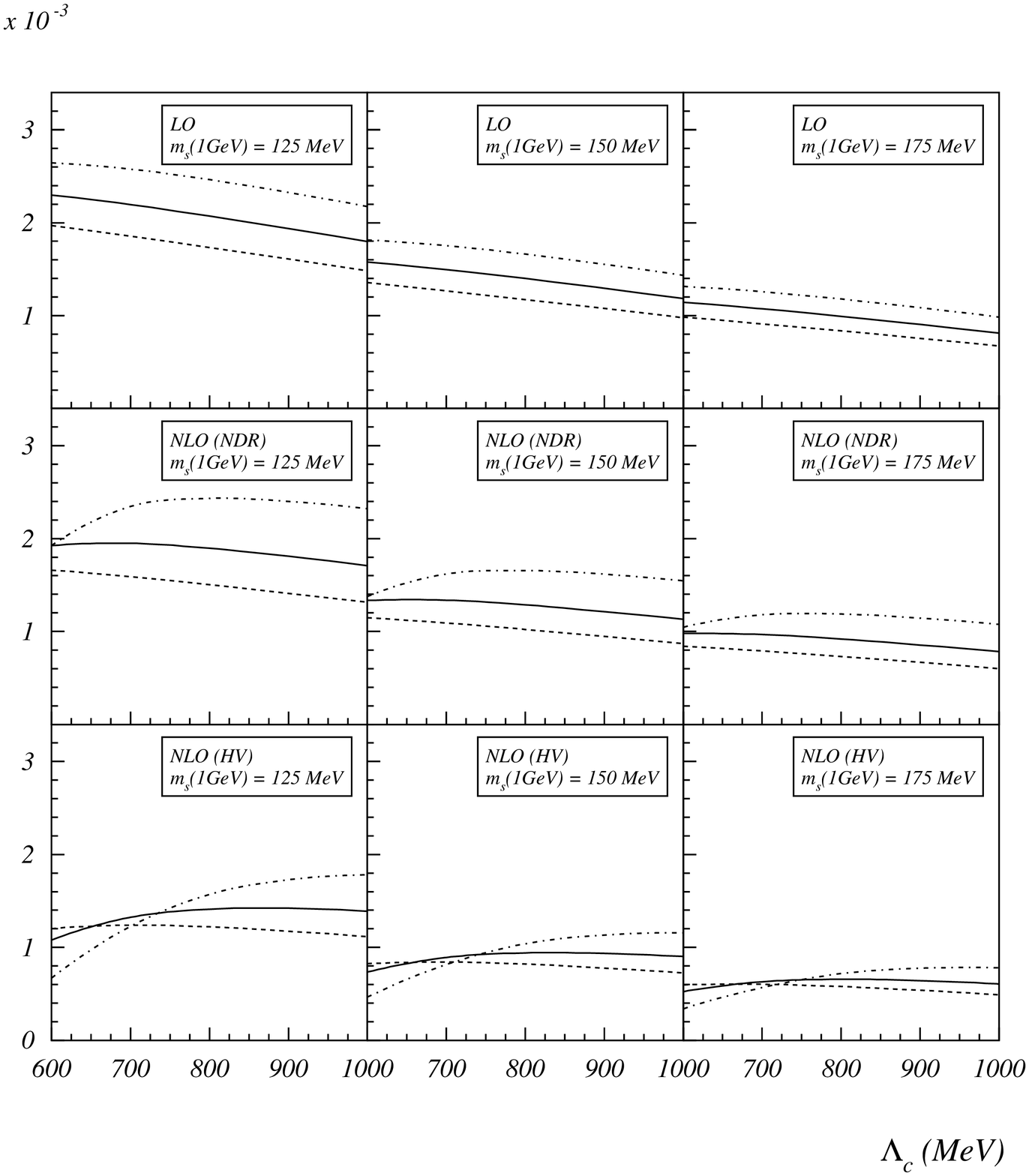,width=17.2cm}}
\caption{$\varepsilon'/\varepsilon$ using LO and NLO (NDR and HV) Wilson
coefficients and the experimental phases, plotted for various values of 
$m_s$ as a function of the matching scale $\lc=\mu$, now with $1/F_\pi
\rightarrow 1/F_K$ in the next-to-leading order terms for the matrix 
elements. We use the central values for $\mbox{Im}\lambda_t$ and 
$\Omega_{\eta+\eta'}$. The solid (dashed, dot-dashed) lines correspond 
to $\Lambda_{\mbox{\tiny QCD}}=325$ $(245,\,405)\,\,\mbox{MeV}$.
\label{epspfk}}
\end{figure}
         
In Fig.~\ref{epspfk} we depict $\varepsilon'/\varepsilon$ as a function of 
the matching scale ($\mu=\Lambda_c$), calculated with LO and NLO (NDR and 
HV) Wilson coefficients and for the central values of $\mbox{Im}\lambda_t$ 
and $\Omega_{\eta+\eta'}$ and various values of $m_s$. The variation of 
$\Omega_{\eta+\eta'}$ does not change the qualitative behaviour; it only 
shifts the curves upward or downward for smaller or larger values of
$\Omega_{\eta+\eta'}$, respectively. The curves in Fig.~\ref{epspfk} result 
from replacing $1/F_\pi$ by $1/F_K$ in all next-to-leading order expressions 
relevant to the complete set of matrix elements. Beside the enhancement of 
the numerical result we also observe a somewhat smaller dependence on the 
matching scale. Finally, even though we obtain somewhat larger values for 
$\varepsilon'/\varepsilon$ the effect is still rather moderate and does not 
affect the statement we made above that lower values of the strange quark 
mass are favoured.

In the above we have argued that estimating the effect of higher order 
corrections to the matrix elements, by replacing the `$1/F_\pi$ expansion'
by a `$1/F_K$ expansion', does not drastically modify the results we 
obtained in the previous section. In particular, this statement refers 
to terms of ${\cal O}(p^0/N_c^2)$ which are corrections on top of the 
${\cal O}(p^0/N_c)$ contribution and correspond to the same pseudoscalar 
representation of the four-quark operator. In the following we will not 
study the $1/N_c^2$ corrections, which correspond to a two-loop diagram
in the chiral theory. The same approximation was made in the chiral quark 
model~\cite{BEFL}. However, estimating the typical effect of higher orders 
by modifying the known corrections ($1/F_\pi\rightarrow 1/F_K$) does not 
account for possible contributions from new terms which are absent at the 
level of the first order corrections. In particular, higher order terms 
in the $p^2$ expansion (tree level) cannot be calculated because the 
low-energy couplings in the ${\cal O}(p^6)$ lagrangian are very uncertain
or even unknown. Nevertheless, these terms are independent of the 
(non-factorizable) matching scale and are chirally suppressed with 
respect to the leading term ($\sim m_{K,\pi}^2/\Lambda_\chi^2$, with 
$\Lambda_\chi\simeq 1\,\mbox{GeV}$ the scale of chiral symmetry breaking).
The convergence of the tree level series was verified for the 
current-current operators~\cite{BBG3,hks} and also for $Q_8$~\cite{HKPSB}, 
where a complete leading plus next-to-leading order calculation now 
exists and tree contributions appear to decrease monotonically. For 
the operator $Q_6$ the leading term is ${\cal O}(p^2)$ and by analogy
we expect the higher order tree terms to be smaller. The terms of 
${\cal O}(p^2/N_c)$ on the other hand are not expected to be small for 
$Q_6$. We remind the readers that for the CP conserving amplitude it is 
mainly the (quadratic) ${\cal O}(p^2/N_c)$ corrections which bring to 
$\langle Q_{1,2}\rangle_0$ a large enhancement relative to the 
(leading-$N_c$) ${\cal O}(p^2)$ values. As the leading-$N_c$ value 
for $Q_6$ is also ${\cal O}(p^2)$ we cannot a priori exclude that the 
value of $\langle Q_6\rangle_0$ is largely affected by ${\cal O}(p^2/N_c)$ 
corrections too. As already discussed in Section~\ref{hme1}, quadratic 
${\cal O}(p^2/N_c)$ corrections are proportional to the factor $\Delta
\equiv\Lambda_c^2/(4 \pi F_\pi)^2$ relative to the ${\cal O}(p^2)$ tree
level contribution.
Different is the case of the operator $Q_8$ since its leading-$N_c$ 
value is ${\cal O}(p^0)$ at lowest order in the chiral expansion. 
Quadratic terms for $Q_8$ are consequently chirally suppressed with respect 
to the leading-$N_c$ value. More precisely the suppression factor is $\sim 
(m_{K,\pi}^2/\Lambda_\chi^2 )\cdot\Delta$. In contrast to $\langle Q_6
\rangle_0$, it is very unlikely that the ${\cal O}(p^2/N_c)$ corrections 
for $\langle Q_8\rangle_2$ could be larger than the ${\cal O}(p^0/N_c)$ 
contributions investigated in the previous section. 

\begin{figure}[t]
\vspace*{2mm}
\centerline{\epsfig{file=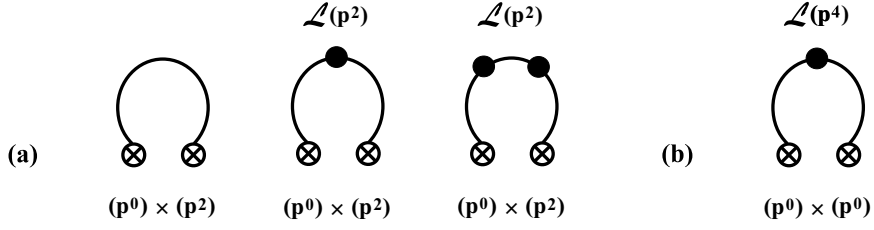,width=11.5cm}}
\caption{Non-factorizable loop diagrams for the evolution of $Q_6$ to 
${\cal O}(p^2/N_c)$. The crossed circles denote the two bosonized densities,
the black circles the strong interaction vertices from the kinetic term
in the lagrangian. Similar diagrams with the ${\cal O}(p^2)$ mass or   
$U_A(1)$ terms in Eq.~(\ref{la2}) are logarithmically divergent or 
finite, and we do not present them here.\label{diagrs}}
\end{figure}

We calculate next the ${\cal O}(p^2/N_c)$ quadratic corrections to the 
matrix elements of the operator $Q_6$. The pseudoscalar representation of 
$Q_6$ can be read off from Eq.~(\ref{dens}):
\begin{eqnarray}
Q_6&=&-2f^2r^2\sum_q \Bigg[ \frac{1}{4}f^2(U^\dagger)_{dq}(U)_{qs} 
+(U^\dagger)_{dq} \big(L_5U\partial_\mu U^\dagger\partial^\mu U 
+2rL_8U{\cal M}^\dagger U \nonumber \\[-2.2mm]
&&+rH_2{\cal M}\big)_{qs}+\big(L_5U^\dagger\partial_\mu U\partial^\mu
U^\dagger+2rL_8U^\dagger{\cal M} U^\dagger+rH_2{\cal M}^\dagger\big)_{dq}
(U)_{qs}\Bigg]+{\cal O}(p^4)\,.\hspace*{6.5mm}\label{q6u}
\end{eqnarray}
In the following we will calculate the ${\cal O}(p^2/N_c)$ evolution
of the operator $Q_6$ in the chiral limit. It is then straightforward  
to compute the hadronic matrix element $\langle Q_6\rangle_0$. To 
calculate the evolution of $Q_6$ we use the background field method as 
described in Ref.~\cite{FG} and also in Refs.~\cite{HKPSB,PSOrsay}. This 
operatorial method is very convenient to calculate corrections in the 
chiral limit. To this end we decompose the matrix $U$ in the classical 
field $\bar{U}$ and the quantum fluctuation $\xi$,
\begin{equation}
U=\exp (i\sqrt{2} \xi/f)\,\bar{U}\;,
\hspace{0.5cm}\xi=\xi^a\frac{\lambda_a}{\sqrt{2}}\,,
\end{equation}
with $\bar{U}$ satisfying the equation of motion
\begin{equation}
\bar{U}\partial^2\bar{U}^\dagger-\partial^2 \bar{U} \bar{U}^\dagger
+r\bar{U}{\cal M}^\dagger-r{\cal M}\bar{U}^\dagger=\frac{\alpha}{N_c}
\langle\ln\bar{U}-\ln\bar{U}^\dagger\rangle\cdot {\bf 1}\;,
\hspace{0.5cm}
\bar{U}=\exp(i\pi^a\lambda_a/f)\;.
\end{equation}
The ${\cal O}(p^2)$ lagrangian thus reads 
\begin{equation}
{\cal L}=\bar{\cal L}
+\frac{1}{2}(\partial_\mu\xi^a\partial^\mu\xi_a)
+\frac{1}{2}\langle [\partial_\mu\xi,\,\xi]\partial^\mu
\bar{U}\bar{U}^\dagger\rangle
-\frac{r}{4}\langle \xi^2\bar{U}{\cal M}^\dagger+\bar{U}^\dagger\xi^2
{\cal M}\rangle-\frac{1}{2}\alpha\xi^0\xi^0+{\cal O}(\xi^3)\;.\label{la2}
\end{equation}
The corresponding expansion of the meson density in Eq.~(\ref{dens}) 
around the classical field yields
\begin{eqnarray}
(D_L)_{ij}&\equiv& \bar{q}_{iR} q_{jL}\,\,=\,\,(\bar{D}_L)_{ij}+irf 
\frac{\sqrt{2}}{4}(\bar{U}^\dagger\xi)_{ji}
+\frac{r}{4}(\bar{U}^\dagger\xi^2)_{ji}\,
\nonumber \\
&&+\,\,i\frac{r}{f} L_5\sqrt{2} \, \big[\,\partial^\mu \bar{U}^\dagger 
\partial_\mu\bar{U} \bar{U}^\dagger \xi +
\bar{U}^\dagger\{\partial^\mu \xi ,\partial_\mu \bar{U} 
\bar{U}^\dagger\}\,\big]_{ji}\,\nonumber \\
&&-\,\,\frac{r}{f^2} L_5 \, \big[\,2 \bar{U}^\dagger 
\partial^\mu \xi \partial_\mu \xi
+\bar{U}^\dagger [ \partial^\mu \xi , \xi ] \partial_\mu \bar{U} 
\bar{U}^\dagger\,\nonumber \\
&&-\,\,2 \bar{U}^\dagger \partial^\mu \xi \partial_\mu \bar{U} \bar{U}^\dagger 
\xi+\partial^\mu \bar{U}^\dagger\{\partial_\mu \xi , \xi \}
-\partial^\mu \bar{U}^\dagger 
\partial_\mu \bar{U} \bar{U}^\dagger \xi^2 \,\big]_{ji}\,
\nonumber \\
&&+\,\,i2 \sqrt{2} \, \frac{r^2}{f} L_8 \,[\,\bar{U}^\dagger \{\xi ,
{\cal M} \bar{U}^\dagger \}\,]_{ji}\,
+2\, \frac{r^2}{f^2} L_8 \,[\,\bar{U}^\dagger \{ \xi^2 , {\cal M} 
\bar{U}^\dagger \}\,]_{ji}\,
+{\cal O}(\xi^3)\,.\hspace*{8mm}\label{Dexp}
\end{eqnarray}
Using Eq.~(\ref{Dexp}) the evolution of $Q_6$ can be obtained in a 
straightforward way. Integrating over the quantum fluctuation by calculating 
the non-factorizable diagrams of Fig.~\ref{diagrs}.a we get the following
result:
\begin{equation}
Q_6(\Lambda_c^2)=-4 F_\pi^2 r^2 \hat{L}_5^r 
(\partial^\mu \bar{U}^\dagger \partial_\mu\bar{U})_{ds}\,
\Bigg[1+\frac{3}{2} \frac{\Lambda_c^2}{(4 \pi F_\pi)^2}\Bigg]\,.
\label{Q6evol}
\end{equation}
This result has already been presented in Ref.~\cite{PSOrsay}. Before 
investigating the numerical effect of the quadratic term in Eq.~(\ref{Q6evol})
a few comments are necessary:
\begin{itemize}
\item In Eq.~(\ref{Q6evol}) we present only the diagonal evolution, 
i.e., the term proportional to the operator $(\partial^\mu \bar{U}^\dagger
\partial_\mu \bar{U})_{ds}$ which gives the only non-vanishing contribution
to the $K\rightarrow\pi\pi$ amplitudes. This property is analogous to the 
tree level. One might note in particular that the $L_8$ contribution 
vanishes since it does not produce a term  proportional to this operator. 
The $H_2$ contribution vanishes from the beginning as it does not 
appear in Eq.~(\ref{Dexp}).
\item To ${\cal O}(p^0/N_c)$ we showed explicitly that the factorizable 
contributions provide the corrections needed to obtain the physical values 
of the low-energy couplings~\cite{HKPSB}. Except for finite corrections, 
the values of the couplings can be obtained in the large-$N_c$ limit, i.e., 
by imposing tree level relations in order to set up the renormalized 
(factorizable) matrix elements [compare Eqs.~(\ref{kp1}) and~(\ref{kp3})]. 
To ${\cal O}(p^2/N_c)$ in Eq.~(\ref{Q6evol}) we use again the renormalized 
coupling $\hat{L}_5^r$, defined in the large-$N_c$ limit\footnote{Note that 
our constants $\hat{L}_i^r$ should not be confused with the renormalization 
scale dependent coefficients $L_i^r$ in Refs.~\cite{GaL} and~\cite{JB}.}, 
since the scale dependence of the bare coefficient $L_5$ will be absorbed 
in factorizable loop corrections to the matrix elements and does not need
to be matched to any short-distance contribution~\cite{HKPSB}.
\item Beside the diagrams in Fig.~\ref{diagrs}.a a priori the diagram
in Fig.~\ref{diagrs}.b with a strong vertex proportional to $L_5$ could 
also contribute. However, it is easy to see that since the $L_5$ term in the 
lagrangian contains a quark mass matrix ${\cal M}$, this diagram produces
an operator similar to the one resulting from the $L_8$ term at tree level 
for $Q_6$. This operator does not contribute to the $K\rightarrow\pi\pi$ 
amplitudes.
\item The diagram of Fig.~\ref{diagrs}.b with the strong vertex 
coming from the $L_1$, $L_2$, $L_3$, and $L_8$ couplings also turns 
out to vanish.
\item There are no ${\cal O}(p^2/N_c)$ tree level contributions,
i.e., from couplings $L_i$ \cite{GaL} which are subleading in $N_c$.
\end{itemize}
\renewcommand{\topfraction}{1.0}
\renewcommand{\textfraction}{0.0}
\noindent

Numerically, we observe a large positive correction from the quadratic
term of ${\cal O}(p^2/N_c)$ in Eq.~(\ref{Q6evol}). The slope of this 
correction is qualitatively consistent and welcome since it compensates 
for the logarithmic decrease at ${\cal O}(p^0/N_c)$. Adding the ${\cal O}
(p^2/N_c)$ term to the full ${\cal O}(p^2)$ and ${\cal O}(p^0/N_c)$
result in Eq.~(\ref{rm3}) we obtain the following matrix element for $Q_6$:
\begin{equation}
\langle Q_6\rangle_0
=-\frac{4\sqrt{3}}{F_\pi}R^2(m_K^2-m_\pi^2)\left[\hat{L}_5^r
\left(1+\frac{3}{2}\frac{\Lambda_c^2}{(4\pi)^2 F_\pi^2}\right)
-\frac{3}{16\,(4\pi)^2}\,\log\Lambda_c^2\,+\,\cdots\,\right].
\hspace*{5mm}\label{q6total}
\end{equation}
\begin{table}[t]
\begin{eqnarray*}
\begin{array}{|c||c|c|c|c|c|c|}\hline
\lc&0.6\,\,\mbox{GeV}&0.7\,\,\mbox{GeV}&
0.8\,\,\mbox{GeV}&0.9\,\,\mbox{GeV}&1.0\,\,\mbox{GeV} \\
\hline\hline
\rule{0cm}{5mm}
B_{6}^{(1/2)}& 1.50 & 1.51 & 1.55 & 1.62 & 1.73 \\[0.5mm]
\hline
\end{array}
\end{eqnarray*}
\caption{$B_{6}^{(1/2)}$, now including, in the chiral limit, terms of 
${\cal O}(p^2/N_c)$.\label{tab8}}
\end{table}
The corresponding values of $B_6^{(1/2)}$ (obtained by adding the quadratic 
corrections to the values in Tab.~\ref{tab3}) are listed in Tab.~\ref{tab8}. 
The $B_6^{(1/2)}$ factor is found to be rather stable around the value 
$B_6^{(1/2)}\simeq 1.6\pm 0.1$. The quadratic term of ${\cal O}(p^2/N_c)$ 
is of the same magnitude as the ${\cal O}(p^2)$ tree term. For the $\langle 
Q_1\rangle_0$ and $\langle Q_2\rangle_0$ there was also a large enhancement 
to the ${\cal O}(p^2)$ tree contribution introduced by the ${\cal O}
(p^2/N_c)$ term (see Tabs.~\ref{tab1} and~\ref{tab3}). $Q_6$ is a 
$\Delta I=1/2$ operator, and the enhancement of $\langle Q_6\rangle_0$ 
suggests that at the level of the $1/N_c$ corrections the dynamics of 
the $\Delta I=1/2$ rule also applies to $Q_6$. One might however note 
that the long-distance evolution of the operator $Q_6$ including both 
the ${\cal O}(p^0/N_c)$ and ${\cal O}(p^2/N_c)$ terms is very different 
from the one of $Q_1$ or $Q_2$. The former is approximately constant 
over a wide range of the cutoff scale due to the smaller coefficient
of the quadratic term and a large cancellation of the scale dependences 
between the quadratic and the logarithmic term, whereas the later [which 
does not receive any ${\cal O}(p^0/N_c)$ contribution] has a large positive 
slope. One should also remark that we observe a noticeable suppression of 
$\langle Q_8\rangle_2$ similar to the one needed for $\langle Q_{1,2}
\rangle_2$ in order to bring the CP conserving $\Delta I=3/2$ amplitude 
down to the experimental value.\footnote{The mechanism for the suppression 
of $\langle Q_8\rangle_2$ however differs from the one for $\langle Q_{1,2}
\rangle_2$, since in the former case it occurs through logarithms and in 
the latter case mainly through quadratic terms \cite{BBG3,hks}.}

\begin{figure}[t]
\vspace*{-1cm}
\centerline{\epsfig{file=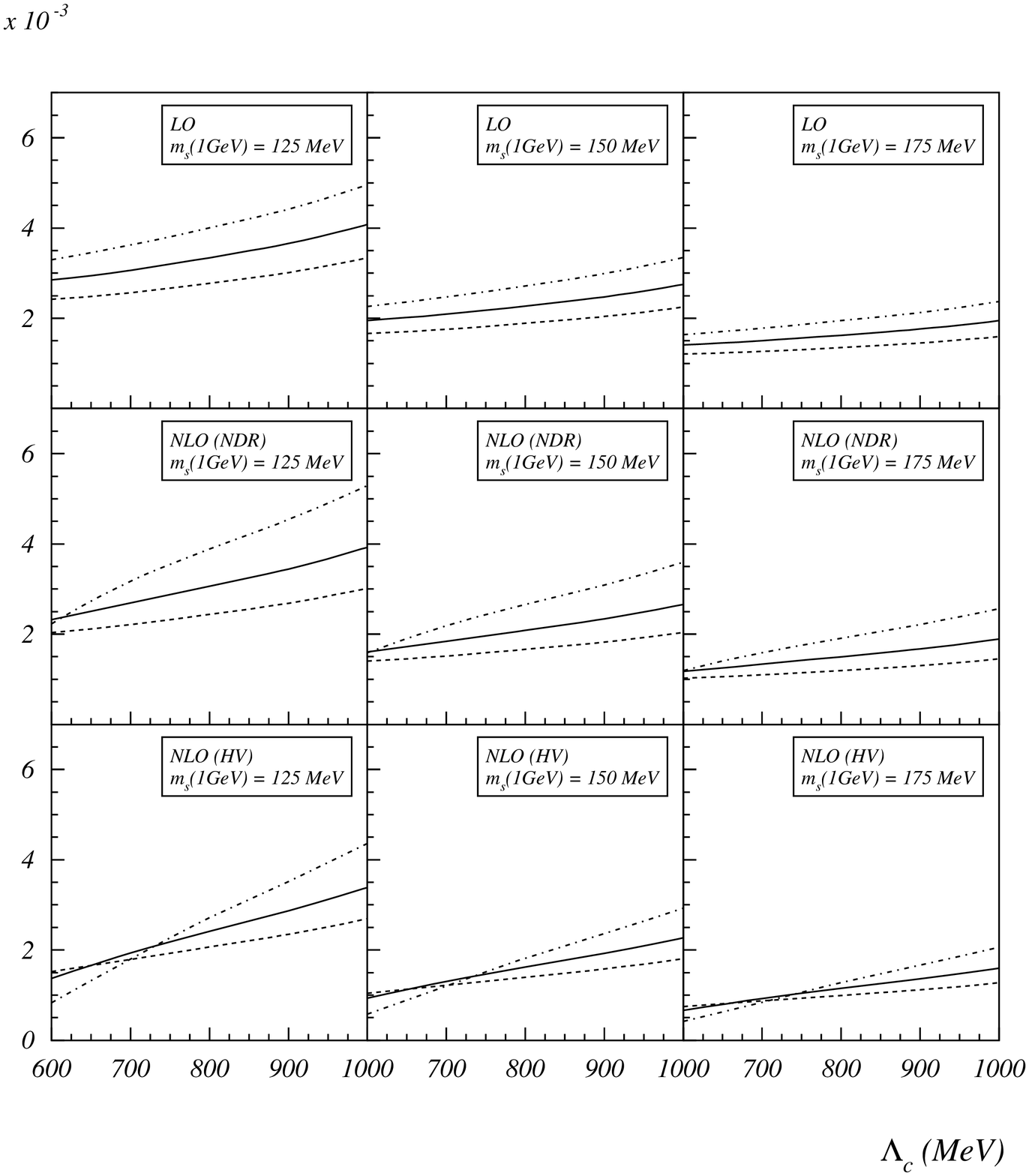,width=17.2cm}}
\caption{$\varepsilon'/\varepsilon$ using LO and NLO (NDR and HV) Wilson 
coefficients and the experimental phases, plotted for various values of
$m_s$ as a function of the matching scale $\lc=\mu$, including for $Q_6$
the term of ${\cal O}(p^2/N_c)$ in the chiral limit. We use again the
values $\mbox{Im}\lambda_t=1.33\cdot 10^{-4}$ and $\Omega_{\eta+\eta'}
=0.25$. The solid (dashed, dot-dashed) lines are for $\Lambda_{\mbox{\tiny 
QCD}}=325$ $(245,\,405)\,\,\mbox{MeV}$.\label{epsnlo}}
\end{figure}

Using the values for $B_6^{(1/2)}$ in Tab.~\ref{tab8} together with 
the bag factors of the remaining operators presented in the previous 
section we calculated again the ratio $\varepsilon'/\varepsilon$ for 
central values of $m_s$, $\Omega_{\eta+\eta'}$, Im$\lambda_t$, and
$\Lambda_{\mbox{\tiny QCD}}$. The results for the three sets of Wilson 
coefficients LO, NDR, and HV and for $\Lambda_c$ between 600 and $900\,
\mbox{MeV}$ are given in Tab.~\ref{tab9}. The numbers are obtained from  
Eqs.~(\ref{Pi0b})\,-\,(\ref{Pi2b}) and (\ref{Pi0b2})\,-\,(\ref{Pi2b2}),
respectively. With LO Wilson coefficients, the enhancement of $B_6^{(1/2)}$
leads to larger values for $\varepsilon'/\varepsilon$, and the predictions 
are now more stable and closer to the data. The results in the NDR scheme
are rather close to the LO ones although more sensitive to the
value of $\Lambda_{\mbox{\tiny QCD}}$. In the HV scheme, the effect of 
the NLO coefficients is more pronounced. The results are significantly
smaller for low values of the matching scale and less stable. This is
illustrated in Fig.~\ref{epsnlo} where we show $\varepsilon'/\varepsilon$ 
for various values of $m_s$ as a function of the matching scale. 

Performing a scanning of the input parameters as explained above, we arrive 
at the values in Tab.~\ref{tab10}. Comparing these results with the values in 
Tabs.~\ref{tab5} and~\ref{tab6} we see a clear enhancement originating from 
the quadratic term in Eq.~(\ref{q6total}). The large ranges reported in 
Tab.~\ref{tab10} can be traced back to the large ranges of the input 
parameters. The parameters, to a large extent, act multiplicatively, and 
the larger range for $\varepsilon'/\varepsilon$ is due to the fact that 
the central value(s) for the ratio are enhanced roughly by a factor of 
two compared to the results we presented in the previous section. More 
accurate information on the parameters, from theory and experiment, will 
restrict the values for the CP ratio.
\begin{table}[t]
\begin{eqnarray*}
\begin{array}{|c||c|c|}\hline
 & \mbox{Case}\,\, 1& \mbox{Case}\,\, 2\\
\hline\hline 
\rule{0cm}{5mm} 
\mbox{LO}& \,\,19.5 \,\,\leq\,\,\varepsilon'/\varepsilon\,\, 
\leq\,\,24.7 \,\, 
& \,\,14.8 \,\,\leq\,\,\varepsilon'/\varepsilon\,\, 
\leq\,\,19.4 \,\, \\[0.5mm]
\mbox{NDR}& \,\,16.1 \,\,\leq\,\,\varepsilon'/\varepsilon\,\, 
\leq\,\,23.4 \,\, 
& \,\,12.5 \,\,\leq\,\,\varepsilon'/\varepsilon\,\,
\leq\,\,18.3 \,\, \\[0.5mm]
\mbox{HV}& \,\,9.3\,\,\,\,\leq\,\,\varepsilon'/\varepsilon\,\, 
\leq\,\,19.3 \,\, 
& \,\,7.0\,\,\,\,\leq\,\,\varepsilon'/\varepsilon\,\,
\leq\,\,14.9 \,\, \\[0.5mm]
\hline
\,\mbox{LO}\,+\,\mbox{NDR}\,+\,\mbox{HV}\,
& \,\,9.3\,\,\,\,\leq\,\,\varepsilon'/\varepsilon\,\,\leq \,\,24.7\,\, 
& \,\,7.0 \,\,\,\,\leq\,\,\varepsilon'/\varepsilon\,\,
\leq\,\,19.4\,\,\\[0.5mm]
\hline
\end{array}
\end{eqnarray*}
\caption{Same results as in Tab.~\ref{tab5}, but now including quadratic 
terms of ${\cal O}(p^2/N_c)$ for $Q_6$ as explained in the text.
\label{tab9}}
\end{table}
\begin{table}[t]
\begin{eqnarray*}
\begin{array}{|c||c|c|}\hline
 & \mbox{Case}\,\, 1& \mbox{Case}\,\, 2\\
\hline\hline 
\rule{0cm}{5mm} 
\mbox{LO}& \,\,8.0 \,\,\leq\,\,\varepsilon'/\varepsilon\,\, 
\leq\,\,62.1 \,\, 
& \,\, 6.1\,\,\leq\,\,\varepsilon'/\varepsilon\,\, 
\leq\,\, 48.5\,\, \\[0.5mm]
\mbox{NDR}& \,\,6.8 \,\,\leq\,\,\varepsilon'/\varepsilon\,\, 
\leq\,\,63.9 \,\, 
& \,\,5.2 \,\,\leq\,\,\varepsilon'/\varepsilon\,\,
\leq\,\,49.8 \,\, \\[0.5mm]
\mbox{HV}& \,\,2.8 \,\,\leq\,\,\varepsilon'/\varepsilon\,\, 
\leq\,\,49.8 \,\, 
& \,\,2.2 \,\,\leq\,\,\varepsilon'/\varepsilon\,\,
\leq\,\,38.5 \,\, \\[0.5mm]
\hline
\,\mbox{LO}\,+\,\mbox{NDR}\,+\,\mbox{HV}\,
& \,\,2.8 \,\,\leq\,\,\varepsilon'/\varepsilon\,\,\leq \,\,63.9\,\, 
& \,\,2.2 \,\,\leq\,\,\varepsilon'/\varepsilon\,\,
\leq\,\,49.8 \,\,\\[0.5mm]
\hline
\end{array}
\end{eqnarray*}
\caption{Same results as in Tab.~\ref{tab6}, but now including quadratic
terms of ${\cal O}(p^2/N_c)$ for $Q_6$ as explained in the text.
\label{tab10}}
\end{table}
\renewcommand{\topfraction}{1.0}
\renewcommand{\textfraction}{0.0}
\noindent

The contributions from current-current operators to $\varepsilon'/
\varepsilon$ are rather small, and corrections from higher order 
terms (beyond the ones investigated in Section~\ref{ana}) and from higher 
resonances will not be able to modify their $B$ factors in a way that 
they change the ratio $\varepsilon'/\varepsilon$ considerably. For the 
reason explained earlier in this article, higher order corrections for 
the operator $Q_8$ are not expected to bring a large change for the 
ratio. The question of whether a large value of $\varepsilon'/\varepsilon$ 
can be accommodated in the standard model without specific values of 
various parameters reduces essentially to the value of $B_6^{(1/2)}$.
Adding the ${\cal O}(p^2/N_c)$ quadratic terms produces a substantial 
increase for the value of the matrix element $\langle Q_6 \rangle_0$, 
and a large value for $B_6^{(1/2)}$ in the range of $1.6$ cannot 
be excluded. This property leads to a more natural explanation for a 
large value of $\varepsilon'/\varepsilon$. Our result can be modified 
by corrections of ${\cal O}(p^2/N_c)$ beyond the chiral limit, from 
logarithms and finite terms, but they are not expected to remove the 
large enhancement observed in Eq.~(\ref{q6total}). Nevertheless it 
would be very interesting to verify this statement through an explicit
calculation.   

In view of the large corrections one might question the convergence of 
the $1/N_c$ expansion. However, we remind the readers that the quadratic 
term of ${\cal O}(p^2/N_c)$ in $Q_6$ we consider in this section 
represents a new class of terms absent to ${\cal O}(p^2)$ and ${\cal O}
(p^0/N_c)$. It is reasonable to assume that the terms of ${\cal O}
(p^2/N_c)$ ($\sim\Delta$) carry a large fraction of the entire
(non-factorizable) contribution, since quadratic corrections in 
the cutoff from higher order terms are chirally suppressed (i.e.,
they are $\sim\Delta\cdot\delta$, $\Delta\cdot\delta^2$, \ldots).

We point out that the quadratic terms obtained at the level of the 
pseudoscalar mesons are physical and must be included in the numerical 
analysis. Our result is compatible with the fact that, in a complete
theory of mesons, the quadratic dependence on the cutoff should be 
absent. Indeed one expects that incorporating higher resonances allows 
one to select higher values for the cutoff and does not remove the 
effect of the quadratic terms, but turns them smoothly into logarithms. 
Therefore, within a limited range of the cutoff, the quadratic terms 
provide an approximate representation of the effect of higher resonances.
This behaviour has been observed in the calculation of the $\pi^+-\pi^0$ 
mass difference after including the vector and axial vector mesons 
\cite{DM,BBiG}. In this particular case the quadratic terms turn 
into finite terms. It has also been observed partly for the 
$\hat{B}_K$ parameter after including the vector mesons~\cite{gmpi}. 
The two examples in Refs.~\cite{BBiG,gmpi} show that effects of higher 
resonances can modify noticeably the results and reduce the dependence 
on the cutoff but are not strong enough to reverse the effects of the 
pseudoscalar mesons. Nevertheless, it would be very interesting to 
include higher resonances in the calculation of the $K\rightarrow
\pi\pi$ decays, in order to study explicitly~their~effect.
  
We note that the $\Delta I =1/2$ enhancement we observe for $Q_6$ 
to ${\cal O}(p^2/N_c)$ is not able to render the contribution 
of $Q_6$ in Re$A_0$ dominant. The numbers for $B_6^{(1/2)}$ in 
Tab.~\ref{tab8} are consistent with the observed value for the $\Delta 
I=1/2$ amplitude, which is dominated by $Q_1$ and $Q_2$; at LO and at NLO 
in the HV scheme, the enhancement of the $B_6^{(1/2)}$ factor changes the 
result of Ref.~\cite{hks} by less than $5\,\%$ for $m_s(1\,\mbox{MeV})=150
\,\mbox{MeV}$ and $245\,\mbox{MeV}\leq\Lambda_{\mbox{\tiny QCD}}\leq 405\,
\mbox{MeV}$. In the NDR scheme, the effect can amount to approximately 
$11\,\%$ of the amplitude for $\Lambda_{\mbox{\tiny QCD}}=405\,\mbox{MeV}$. 
Therefore we do not see a correlation between the large values of Re$A_0$ 
and $\varepsilon'/\varepsilon$, since the two quantities are dominated by 
different operators.\footnote{For a discussion of this point see also 
Ref.~\cite{bosch}.} In particular, in one of the first estimates of 
$\varepsilon'/\varepsilon$ \cite{gwcp} it was suggested (following 
Ref.~\cite{VSZ}) that the $\Delta I=1/2$ amplitude is dominated by the 
operator $Q_6$, which would lead to a large value of $\varepsilon'/
\varepsilon$. Such a mechanism, in the range for the matching scale 
we consider, would require an enhancement of $\langle Q_6 \rangle_0$ 
several times larger than the one obtained in the present analysis. 
The same remark applies to~Ref.~\cite{kns}.

Among the previous calculations, loop corrections to the operators 
$Q_6$ and $Q_8$ in the $1/N_c$ expansion were also considered in 
Refs.~\cite{china,DO1}. This study used a different matching condition, 
and the parametrization of the ${\cal O}(p^4)$ lagrangian was not general. 
The authors obtained a large reduction of $\langle Q_8\rangle_2$ and an 
enhancement of $\langle Q_6\rangle_0$, predicting large values for 
$\varepsilon'/\varepsilon$. 

It is interesting to compare our values for $\varepsilon'/\varepsilon$ 
with the results found with other methods. Lattice calculations obtained
$B_6^{(1/2)}(2\,\mbox{GeV})=1.0\pm 0.2$ and $B_8^{(3/2)}(2\,\mbox{GeV})=1.0
\pm 0.2$ and predicted a small value for $\varepsilon'/\varepsilon=(4.6\pm 
3.0)\cdot 10^{-4}$ with Gaussian errors for the experimental input (see 
Ref.~\cite{epsrome} and references therein). More recent values reported 
for $B_8^{(3/2)}$ are $B_8^{(3/2)}(2\,\mbox{GeV})=0.81(3)(3)$~\cite{BGS}, 
0.77(4)(4)~\cite{kgs}, and 1.03(3)~\cite{Conti}. $B_6^{(1/2)}$ was 
estimated in Ref.~\cite{peki1}: $B_6^{(1/2)}(2\,\mbox{GeV})=0.76(3)(5)$.
However, as stressed in Ref.~\cite{gupta}, the systematic uncertainties
in this calculation are not completely under control. This statement has 
been confirmed by a recent analysis~\cite{peki2} which obtained negative
values for $B_6^{(1/2)}$ and favours either negative or slightly positive 
values for $\varepsilon'/\varepsilon$. Although the scales used in lattice 
calculations and the phenomenological approaches are different, the 
various results for the $B$ factors can be compared, for values of the
scale around $1\,\mbox{GeV}$ or above, since $B_6^{(1/2)}$ and 
$B_8^{(3/2)}$ were shown in QCD to depend only very weakly on the 
renormalization scale for values above $1\,\mbox{GeV}$~\cite{BJM}. 
Small values for $\varepsilon'/\varepsilon$ consistent with zero
were also quoted in~Ref.~\cite{belkov}.

The chiral quark model \cite{BEFL} yields a range for $B_6^{(1/2)}$ in the 
HV scheme which is above the VSA value, $B_6^{(1/2)}(0.8\,\mbox{GeV})=1.6
\pm 0.3$, and predicts a small reduction of the $B_8^{(3/2)}$ factor, 
$B_8^{(3/2)}(0.8\,\mbox{GeV})=0.92\pm 0.02$. The quoted range for the CP 
ratio is $7\cdot 10^{-4}\,\leq\,\varepsilon'/\varepsilon\,\leq\,31\cdot 
10^{-4}$. Since the treatment of the renormalization scale in 
Ref.~\cite{BEFL} is different from the one used in this article we 
do not see a clear link which could easily explain why both approaches 
give approximately the same result for $B_6^{(1/2)}$. 

Very recently, an extensive study of $\varepsilon'/\varepsilon$ in the 
standard model was presented in Ref.~\cite{bosch}. The authors investigated 
the sensitivity of the CP ratio on the input parameters and updated their 
numerical values. They treated $B_6^{(1/2)}$ and $B_8^{(3/2)}$ as parameters
and adopted the values $B_6^{(1/2)}=1.0\pm 0.3$ and $B_8^{(3/2)}=0.8\pm 0.2$ 
together with the constraint $B_6^{(1/2)}\geq B_8^{(3/2)}$. Numerically, 
they obtained $1.05\cdot 10^{-4}\,\leq\,\varepsilon'/\varepsilon\,\leq\,
28.8\cdot 10^{-4}$ and $0.26\cdot 10^{-4}\,\leq\,\varepsilon'/\varepsilon
\,\leq\,22.0\cdot 10^{-4}$ in the NDR and HV schemes, respectively. The 
quoted results are consistent with the values we get for $\varepsilon'/
\varepsilon$ to ${\cal O}(p^2)$ and ${\cal O}(p^0/N_c)$ for $Q_6$ and 
$Q_8$. As shown in this section, an additional large contribution comes
from the ${\cal O}(p^2/N_c)$ term of the $Q_6$ operator.  

In summary, we have shown in this section that the quadratic terms 
of ${\cal O}(p^2/N_c)$ are large for $Q_6$. From general counting arguments
we have good indications that among the various next-to-leading order terms 
in the $p^2$ and $1/N_c$ expansions they are the dominant ones. They enhance 
$B_6^{(1/2)}$ and bring $\varepsilon'/\varepsilon$ much closer to the 
measured value for central values of the input parameters. We obtain a 
quadratic evolution for $Q_6$ which indicates that a $\Delta I=1/2$ 
enhancement mechanism is operative for $Q_6$ as for $Q_{1,2}$. 
$B_8^{(3/2)}$ is expected to be affected much less by terms of 
${\cal O}(p^2/N_c)$ due to an extra $p^2$ suppression factor 
relative to the leading ${\cal O}(p^0)$ tree term. 

One should recall that our analysis is performed in the chiral limit. 
Corrections beyond the chiral limit, from logarithms and finite terms, are 
not expected to remove the large enhancement of $B_6^{(1/2)}$ arising from 
the quadratic term in Eq.~(\ref{q6total}).\footnote{Note that for the CP 
conserving $\Delta I=1/2$ amplitude the chiral limit gives a good 
approximation to the numerical result.} Consequently, the results for 
$\varepsilon'/\varepsilon$ we obtained in Section~\ref{ana}, by including 
terms of ${\cal O}(p^2)$ and ${\cal O}(p^0/N_c)$ for $Q_6$ and $Q_8$, should 
be considered as a lower range, which is shifted to higher values by including 
also quadratic terms of ${\cal O}(p^2/N_c)$. The ideal case would be to 
calculate and include the full ${\cal O}(p^2/N_c)$ amplitudes, as well as 
the ${\cal O}(p^4)$ and ${\cal O}(p^0/N_c^2)$ terms. It would also be
interesting to investigate the effect of higher resonances (at least the 
vector mesons and presumably also the axial vector and scalar mesons). 
Each of the additional effects separately is not expected to counteract 
largely the enhancement found for $B_6^{(1/2)}$. Nevertheless, in the 
extreme (and unlikely) case where all these effects would come with 
the same sign a significant modification of the result cannot be 
excluded formally. In order to reduce the scheme dependence in the 
result for $\varepsilon'/\varepsilon$ appropriate subtractions would 
be necessary (see Refs.~\cite{BPdelta,BB}).

In view of the noticeable uncertainties connected still with both the 
calculation of the matrix elements and the exact values of the various 
parameters taken from theory and experiment, it is difficult to decide
whether the large value of $\varepsilon'/\varepsilon$ observed recently
is indicating new physics beyond the standard model~\cite{kns,sil,mm}. In 
this situation it is interesting to investigate other kaon decays in order 
to perform precision tests of flavour dynamics and to search for new 
physics~\cite{ab98,isi}.
%
%
\section{Conclusions \label{con}}
In this article we have presented results of a new analysis for the 
CP parameter $\varepsilon'/\varepsilon$. Our interest in this topic 
concentrates on the improved calculation of loop corrections for the 
hadronic matrix elements. It is well known that the leading-$N_c$ values 
for the matrix elements underestimate the $\Delta I=1/2$ amplitude $A_0$ 
in $K\to\pi\pi$ decays. It has been shown earlier that $1/N_c$ contributions 
to the operators $Q_1$ and $Q_2$ are very large, bringing the value for 
the amplitude $A_0$ closer to the experimental value~\cite{BBG3}; an 
improved matching condition brings it even closer~\cite{hks}. The same 
method introduced corrections to the matrix elements of the operators 
$Q_6$ and $Q_8$ \cite{HKPSB,china,DO1} and modified the predictions for 
the parameter $\varepsilon'/\varepsilon$.

In view of this knowledge and the fact that three large experiments 
\cite{barr,gibb,ktev} were measuring the CP asymmetry, we decided to 
embark on an extensive study of the hadronic matrix elements including 
at the same time improvements of the input parameters which have taken 
place in the meantime. In particular, we incorporate an improved estimate 
of the multiplicative CKM factor~\cite{bosch} and use leading and 
next-to-leading order Wilson coefficients, which were communicated 
to us by M. Jamin~\cite{jam}.

In the first part of the paper (up to Section 4) we have presented our
results to ${\cal O}(p^2)$ and ${\cal O}(p^0/N_c)$ for the dominant 
operators $Q_6$ and $Q_8$~\cite{HKPSB} and have included them in an 
extensive analysis of the CP parameter. We have found that the matrix 
elements $\langle Q_6\rangle_0$ and $\langle Q_8\rangle_2$, to this order, 
have only a logarithmic dependence on the cutoff. The corrections to these 
operators are smaller than those of $\langle Q_1\rangle_0$ and $\langle 
Q_2\rangle_0$ which are quadratic in the cutoff~\cite{hks}. They decrease 
$\langle Q_8\rangle_2$ roughly to half its value in the VSA and modify 
$\langle Q_6\rangle_0$ to a lesser extent leading to a ratio $B_6^{(1/2)}
/B_8^{(3/2)}\simeq 1.8$. The net effect is to eliminate the almost complete 
cancellation between the two operators but the overall values of the matrix
elements are reduced. The corresponding ranges for $\varepsilon'/
\varepsilon$ are given in Tabs.~\ref{tab5}\,-\,\ref{tab6} and 
Figs.~\ref{epsp1}\,-\,\ref{epsp3}. Adopting central values for 
the input parameters ($m_s$, $\Omega_{\eta+\eta'}$, 
$\Lambda^{(4)}_{\overline{\mbox{\tiny MS}}}$, and Im$\lambda_t$) 
we obtain $4.2\cdot 10^{-4}\,\leq\,\varepsilon'/\varepsilon\,
(\mbox{central})\,\leq\,12.9\cdot 10^{-4}$. The quoted range refers 
to the uncertainties associated with the calculation of the hadronic 
matrix elements and with the use of three sets of Wilson coefficients 
LO, NDR, and HV. Performing a complete scanning of the parameter space 
we obtain
\begin{displaymath}
1.5\cdot 10^{-4}\,\,\leq\,\,\varepsilon'/\varepsilon
\,\,\leq\,\,31.6\cdot 10^{-4}\,.
\end{displaymath}
The upper values for $\varepsilon'/\varepsilon$ obtained in this part of the 
article are close to the experimental data \cite{barr,ktev}. They are reached 
only for low values of $m_s$ and specific values for the other parameters. As 
stated in the article, this is the complete first-order calculation for $Q_6$ 
in the twofold expansion and provides a benchmark for additional corrections. 

A major part of the present article is the estimate of still higher
order effects. In this direction we have studied, first, the changes 
introduced by the replacement of the coupling constant $F_\pi$ by $F_K$ 
in the next-to-leading order expressions for the matrix elements, which 
gives an indirect estimate of higher order corrections~\cite{HKPSB}. We 
found that the predicted values are increased. Numerical results for 
central values of $\mbox{Im}\lambda_t$ and $\Omega_{\eta+\eta'}$ and 
various values of $m_s$ are shown in Fig.~\ref{epspfk}, which indicate 
that the experimental data can be accommodated in the standard model. 
A low value of $m_s$ is also favoured.

In a second step we studied the ${\cal{O}}(p^2/N_c)$ corrections for 
$Q_6$. Here the ${\cal O}(p^0)$ term vanishes; the ${\cal O}(p^0/N_c)$ 
correction was found to be moderate~\cite{HKPSB}. Thus a significant 
correction appears, for the first time, through quadratic terms of 
${\cal O}(p^2/N_c)$, and the behaviour of $\langle Q_6\rangle_0$ is 
similar to the one found for the matrix elements $\langle Q_1\rangle_0$ 
and $\langle Q_2\rangle_0$~\cite{hks}. In Section~5 we have shown that the 
value for $\langle Q_6\rangle_0$ is enhanced by the ${\cal O}(p^2/N_c)$ 
contribution in the chiral limit. This point we already emphasized in 
Ref.~\cite{PSOrsay}. Numerically, we obtain values for $B_6^{(1/2)}$ 
around $1.6\pm 0.1$. Our calculation indicates that at the level of the 
$1/N_c$ corrections a $\Delta I=1/2$ enhancement is operative for $Q_6$ 
similar to the one of $Q_1$ and $Q_2$ which dominate the CP conserving 
amplitude. The effect of adding the ${\cal O}(p^2/N_c)$ quadratic terms 
is evident as a substantial increase in the value of $\varepsilon'/
\varepsilon$, which brings the result rather close to the data for central 
values of the input parameters. Numerically, this is shown by the following 
range obtained by collecting together the results for the three sets of 
Wilson coefficients LO, NDR, and HV:
\begin{displaymath}
7.0\cdot 10^{-4}\,\,\leq\,\,\varepsilon'/\varepsilon
\,(\mbox{central})\,\,\leq\,24.7\cdot 10^{-4} 
\end{displaymath}
(for details see Tab.~\ref{tab9}). Performing a complete scanning of the 
parameter space for the various cases produces the ranges reported in 
Tab.~\ref{tab10}. 

As stated in the article, it is still desirable to calculate and compare 
the full amplitudes to ${\cal O}(p^4)$, ${\cal O}(p^2/N_c)$, and ${\cal O}
(p^0/N_c^2)$. The incorporation of higher resonances would be very 
interesting since it would allow to select higher values for the matching 
scale. A more sophisticated treatment of the scheme dependence remains a
challenge for future studies. However, it is encouraging that the 
approximations we made in this paper give results close to the experimental 
data. Clearly the possibility of an natural explanation, within the 
standard model, of the experimental value for $\varepsilon'/\varepsilon$ 
cannot be excluded.

To sum up, we have presented our results from an extensive study of the 
hadronic matrix elements to ${\cal O}(p^2/N_c)$. We computed all matrix 
elements in the same theoretical framework, except for $\langle Q_1
\rangle_2=\langle Q_2\rangle_2$ which were extracted from the data on 
the CP conserving decays. Our predictions for $\varepsilon'/\varepsilon$ 
are close to the weighted experimental average for central values of the 
input parameters.\\[0.6cm] 
\underline{Note added:} After completion of this article the NA48 
collaboration at CERN reported the value $\mbox{Re}\,(\varepsilon'/
\varepsilon)=(18.5\pm 7.3)\cdot10^{-4}$~\cite{sozzi}. The new world
average is $\mbox{Re}\,(\varepsilon'/\varepsilon)=(21.2\pm 4.6)
\cdot10^{-4}$. The conclusions of the present article are in 
agreement with this new measurement. 

\newpage
\renewcommand{\textfraction}{1.0}
\vspace{1.5cm}
\begin{center}{\large Acknowledgements}
\end{center}  
We wish to thank Johan Bijnens, Andrzej Buras, Jorge Fatelo, 
Jean-Marc G\'erard, and Gino Isidori for discussions; especially Bill
Bardeen for helpful advice and discussions throughout this work. We are 
very thankful to Matthias Jamin for providing us with the numerical values 
of the Wilson coefficients used in this article. This work was supported in 
part by the Bundesministerium f\"ur Bildung, Wissenschaft, Forschung und 
Technologie (BMBF), 057D093P(7), Bonn, FRG, and DFG Antrag PA-10-1. One of 
us (T.H.) acknowledges partial support from EEC, TMR-CT980169.
%

\newpage
\begin{appendix}
\section{Numerical Values of the Wilson Coefficients \label{append}}
In this appendix we list the numerical values of the LO and NLO 
(HV and NDR) Wilson coefficients for $\Delta S=1$ transitions used in 
Section~\ref{numdis1}. These values were communicated to us by M.~Jamin 
\cite{jam}. Following the lines of Ref.~\cite{BJM} the coefficients $y_i$ 
are given for a 10-dimensional operator basis $\{Q_1,\ldots,Q_{10}\}$. 
Below the charm threshold the set of operators reduces to seven linearly 
independent operators [see Eqs.~(\ref{qia})\,-\,(\ref{qio})] with
\begin{eqnarray}
Q_4\,=\,-Q_1+Q_2+Q_3\,,\hspace*{8mm}
Q_9\,=\,\frac{3}{2}Q_1-\frac{1}{2}Q_3\,,\hspace*{8mm} 
Q_{10}\,=\,\frac{1}{2}Q_1+Q_2-\frac{1}{2}Q_3\,. \label{linear-op} 
\end{eqnarray}
At next-to-leading logarithmic order in (renormalization group improved) 
perturbation theory in the NDR scheme the relations in Eq.~(\ref{linear-op}) 
receive ${\cal O}(\alpha_s)$ and ${\cal O}(\alpha)$ corrections \cite{BJM,BBL}. 
In the present analysis we use the linear dependence at the level of 
the matrix elements $\langle Q_i\rangle_I$, i.e., at the level of the 
pseudoscalar representation where modifications to the relations in 
Eq.~(\ref{linear-op}) are absent. We note that the effect of the different 
treatment of the operator relations at next-to-leading logarithmic order, 
which is due to the fact that in the long-distance part there is no 
(perturbative) counting in $\alpha_s$, is numerically negligible.

The following parameters are used for the calculation of the Wilson
coefficients:
\[
M_W\,=\,80.2\,\mbox{GeV},\hspace*{8mm}\sin^2\theta_W\,=\,0.23,\hspace*{8mm}
\alpha=1/129,
\]
\[
m_t\,=\,170\,\mbox{GeV},\hspace*{8mm}\overline{m}_b(m_b)\,=\,4.4\,\mbox{GeV},
\hspace*{8mm}\overline{m}_c(m_c)\,= \,1.3\,\mbox{GeV}\,.
\]
\vspace*{2mm}
\begin{eqnarray*}
\begin{array}{|c||c|c|c|c|c|c|}\hline
\mu&0.6\,\,\mbox{GeV}&0.7\,\,\mbox{GeV}&0.8\,\,\mbox{GeV}&
0.9\,\,\mbox{GeV}&1.0\,\,\mbox{GeV} \\ 
\hline\hline 
y_{1} & 0.0 & 0.0 & 0.0 & 0.0 & 0.0 \\[0.3mm]
y_{2} & 0.0 & 0.0 & 0.0 & 0.0 & 0.0 \\[0.3mm]
y_{3} & 0.0410 & 0.038 & 0.035 & 0.034 & 0.032 \\[0.3mm]
y_{4} & -0.056 & -0.056 & -0.055 & -0.055 & -0.055 \\[0.3mm]
y_{5} & 0.009 & 0.011 & 0.012 & 0.012 & 0.013 \\[0.3mm]
y_{6} & -0.133 & -0.116 & -0.106 & -0.098 & -0.092 \\[0.3mm]
y_{7}/\alpha & 0.024 & 0.025 & 0.027 & 0.028 & 0.029 \\[0.3mm]
y_{8}/\alpha & 0.217 & 0.180 & 0.155 & 0.138 & 0.125 \\[0.3mm]
y_{9}/\alpha & -1.749 & -1.657 & -1.595 & -1.550 & -1.515 \\[0.3mm]
y_{10}/\alpha & 1.007 & 0.887 & 0.803 & 0.740 & 0.690 \\[0.3mm]
\hline
\end{array}
\end{eqnarray*}\\[2mm]\noindent
\centerline{Table~11:
$\Delta S = 1$ LO Wilson coefficients for
$\Lambda_{\mbox{\tiny QCD}}=245\,\mbox{MeV}$.}
\newpage
\begin{eqnarray*}
\begin{array}{|c||c|c|c|c|c|c|}\hline
\mu&0.6\,\,\mbox{GeV}&0.7\,\,\mbox{GeV}&0.8\,\,\mbox{GeV}&
0.9\,\,\mbox{GeV}&1.0\,\,\mbox{GeV} \\ 
\hline\hline 
y_{1} & 0.0 & 0.0 & 0.0 & 0.0 & 0.0 \\[0.3mm]
y_{2} & 0.0 & 0.0 & 0.0 & 0.0 & 0.0 \\[0.3mm]
y_{3} & 0.052 & 0.046 & 0.043 & 0.040 & 0.038 \\[0.3mm]
y_{4} & -0.063 & -0.063 & -0.062 & -0.062 & -0.062 \\[0.3mm]
y_{5} & 0.008 & 0.010 & 0.012 & 0.013 & 0.013 \\[0.3mm]
y_{6} & -0.187 & -0.154 & -0.135 & -0.122 & -0.113 \\[0.3mm]
y_{7}/\alpha & 0.029 & 0.031 & 0.033 & 0.034 & 0.036 \\[0.3mm]
y_{8}/\alpha & 0.324 & 0.249 & 0.206 & 0.178 & 0.158 \\[0.3mm]
y_{9}/\alpha & -1.957 & -1.799 & -1.702 & -1.634 & -1.585 \\[0.3mm]
y_{10}/\alpha & 1.280 & 1.082 & 0.956 & 0.867 & 0.800 \\[0.3mm]
\hline
\end{array}
\end{eqnarray*}\\[2mm]\noindent
\centerline{Table~12:
$\Delta S = 1$ LO Wilson coefficients for
$\Lambda_{\mbox{\tiny QCD}}=325\,\mbox{MeV}$.}
\vspace*{1.5cm}
\begin{eqnarray*}
\begin{array}{|c||c|c|c|c|c|c|}\hline
\mu&0.6\,\,\mbox{GeV}&0.7\,\,\mbox{GeV}&0.8\,\,\mbox{GeV}&
0.9\,\,\mbox{GeV}&1.0\,\,\mbox{GeV} \\ 
\hline\hline 
y_{1} & 0.0 & 0.0 & 0.0 & 0.0 & 0.0 \\[0.3mm]
y_{2} & 0.0 & 0.0 & 0.0 & 0.0 & 0.0 \\[0.3mm]
y_{3} & 0.065 & 0.057 & 0.051 & 0.048 & 0.045 \\[0.3mm]
y_{4} & -0.069 & -0.070 & -0.069 & -0.069 & -0.069 \\[0.3mm]
y_{5} & 0.005 & 0.009 & 0.011 & 0.013 & 0.014 \\[0.3mm]
y_{6} & -0.285 & -0.209 & -0.173 & -0.152 & -0.137 \\[0.3mm]
y_{7}/\alpha & 0.033 & 0.035 & 0.038 & 0.039 & 0.041 \\[0.3mm]
y_{8}/\alpha & 0.526 & 0.356 & 0.277 & 0.230 & 0.198 \\[0.3mm]
y_{9}/\alpha & -2.295 & -1.995 & -1.836 & -1.736 & -1.666 \\[0.3mm]
y_{10}/\alpha & 1.690 & 1.334 & 1.139 & 1.011 & 0.920 \\[0.3mm]
\hline
\end{array}
\end{eqnarray*}\\[2mm]\noindent
\centerline{Table~13:
$\Delta S = 1$ LO Wilson coefficients for
$\Lambda_{\mbox{\tiny QCD}}=405\,\mbox{MeV}$.}
\newpage
\begin{eqnarray*}
\begin{array}{|c||c|c|c|c|c|c|}\hline
\mu&0.6\,\,\mbox{GeV}&0.7\,\,\mbox{GeV}&0.8\,\,\mbox{GeV}&
0.9\,\,\mbox{GeV}&1.0\,\,\mbox{GeV} \\ 
\hline\hline 
y_{1} & 0.0 & 0.0 & 0.0 & 0.0 & 0.0 \\[0.3mm]
y_{2} & 0.0 & 0.0 & 0.0 & 0.0 & 0.0 \\[0.3mm]
y_{3} & 0.033 & 0.031 & 0.029 & 0.028 & 0.027 \\[0.3mm]
y_{4} & -0.050 & -0.051 & -0.051 & -0.051 & -0.050 \\[0.3mm]
y_{5} & -0.014 & -0.005 & -0.001 & 0.002 & 0.004 \\[0.3mm]
y_{6} & -0.154 & -0.120 & -0.103 & -0.093 & -0.085 \\[0.3mm]
y_{7}/\alpha & -0.037 & -0.035 & -0.034 & -0.033 & -0.032 \\[0.3mm]
y_{8}/\alpha & 0.234 & 0.189 & 0.163 & 0.146 & 0.134 \\[0.3mm]
y_{9}/\alpha & -1.783 & -1.658 & -1.586 & -1.537 & -1.502 \\[0.3mm]
y_{10}/\alpha & 0.971 & 0.803 & 0.700 & 0.630 & 0.577 \\[0.3mm]
\hline
\end{array}
\end{eqnarray*}\\[2mm]\noindent
\centerline{Table~14:
$\Delta S = 1$ NLO Wilson coefficients (NDR) for
$\Lambda_{\mbox{\tiny QCD}}=\Lambda^{(4)}_{\overline{\mbox{\tiny MS}}}
=245\,\mbox{MeV}$.}
\vspace*{1.5cm}
\begin{eqnarray*}
\begin{array}{|c||c|c|c|c|c|c|}\hline
\mu&0.6\,\,\mbox{GeV}&0.7\,\,\mbox{GeV}&0.8\,\,\mbox{GeV}&
0.9\,\,\mbox{GeV}&1.0\,\,\mbox{GeV} \\ 
\hline\hline 
y_{1} & 0.0 & 0.0 & 0.0 & 0.0 & 0.0 \\[0.3mm]
y_{2} & 0.0 & 0.0 & 0.0 & 0.0 & 0.0 \\[0.3mm]
y_{3} & 0.040 & 0.036 & 0.034 & 0.032 & 0.030 \\[0.3mm]
y_{4} & -0.055 & -0.054 & -0.053 & -0.053 & -0.053 \\[0.3mm]
y_{5} & 0.017 & 0.015 & 0.014 & 0.014 & 0.014 \\[0.3mm]
y_{6} & -0.125 & -0.103 & -0.090 & -0.082 & -0.077 \\[0.3mm]
y_{7}/\alpha & -0.033 & -0.033 & -0.033 & -0.032 & -0.032 \\[0.3mm]
y_{8}/\alpha & 0.269 & 0.212 & 0.181 & 0.160 & 0.146 \\[0.3mm]
y_{9}/\alpha & -1.791 & -1.663 & -1.588 & -1.539 & -1.503 \\[0.3mm]
y_{10}/\alpha & 0.990 & 0.816 & 0.711 & 0.638 & 0.585 \\[0.3mm]
\hline
\end{array}
\end{eqnarray*}\\[2mm]\noindent
\centerline{Table~15:
$\Delta S = 1$ NLO Wilson coefficients (HV) for
$\Lambda_{\mbox{\tiny QCD}}=\Lambda^{(4)}_{\overline{\mbox{\tiny MS}}}
=245\,\mbox{MeV}$.}
\newpage
\begin{eqnarray*}
\begin{array}{|c||c|c|c|c|c|c|}\hline
\mu&0.6\,\,\mbox{GeV}&0.7\,\,\mbox{GeV}&0.8\,\,\mbox{GeV}&
0.9\,\,\mbox{GeV}&1.0\,\,\mbox{GeV} \\ 
\hline\hline 
y_{1} & 0.0 & 0.0 & 0.0 & 0.0 & 0.0 \\[0.3mm]
y_{2} & 0.0 & 0.0 & 0.0 & 0.0 & 0.0 \\[0.3mm]
y_{3} & 0.037 & 0.038 & 0.036 & 0.034 & 0.032 \\[0.3mm]
y_{4} & -0.051 & -0.056 & -0.057 & -0.058 & -0.058 \\[0.3mm]
y_{5} & -0.067 & -0.024 & -0.011 & -0.004 & -0.001 \\[0.3mm]
y_{6} & -0.334 & -0.199 & -0.150 & -0.126 & -0.111 \\[0.3mm]
y_{7}/\alpha & -0.052 & -0.037 & -0.034 & -0.032 & -0.031 \\[0.3mm]
y_{8}/\alpha & 0.413 & 0.289 & 0.229 & 0.195 & 0.173 \\[0.3mm]
y_{9}/\alpha & -2.160 & -1.864 & -1.718 & -1.633 & -1.576 \\[0.3mm]
y_{10}/\alpha & 1.445 & 1.079 & 0.889 & 0.771 & 0.690 \\[0.3mm]
\hline
\end{array}
\end{eqnarray*}\\[2mm]\noindent
\centerline{Table~16:
$\Delta S = 1$ NLO Wilson coefficients (NDR) for
$\Lambda_{\mbox{\tiny QCD}}=\Lambda^{(4)}_{\overline{\mbox{\tiny MS}}}
=325\,\mbox{MeV}$.}
\vspace*{1.5cm}
\begin{eqnarray*}
\begin{array}{|c||c|c|c|c|c|c|}\hline
\mu&0.6\,\,\mbox{GeV}&0.7\,\,\mbox{GeV}&0.8\,\,\mbox{GeV}&
0.9\,\,\mbox{GeV}&1.0\,\,\mbox{GeV} \\ 
\hline\hline 
y_{1} & 0.0 & 0.0 & 0.0 & 0.0 & 0.0 \\[0.3mm]
y_{2} & 0.0 & 0.0 & 0.0 & 0.0 & 0.0 \\[0.3mm]
y_{3} & 0.052 & 0.048 & 0.043 & 0.040 & 0.037 \\[0.3mm]
y_{4} & -0.064 & -0.063 & -0.062 & -0.061 & -0.061 \\[0.3mm]
y_{5} & 0.037 & 0.021 & 0.018 & 0.017 & 0.016 \\[0.3mm]
y_{6} & -0.229 & -0.155 & -0.124 & -0.107 & -0.097 \\[0.3mm]
y_{7}/\alpha & -0.023 & -0.030 & -0.031 & -0.031 & -0.030 \\[0.3mm]
y_{8}/\alpha & 0.500 & 0.329 & 0.255 & 0.214 & 0.188 \\[0.3mm]
y_{9}/\alpha & -2.188 & -1.875 & -1.724 & -1.636 & -1.577 \\[0.3mm]
y_{10}/\alpha & 1.489 & 1.102 & 0.904 & 0.783 & 0.699 \\[0.3mm]
\hline
\end{array}
\end{eqnarray*}\\[2mm]\noindent
\centerline{Table~17:
$\Delta S = 1$ NLO Wilson coefficients (HV) for
$\Lambda_{\mbox{\tiny QCD}}=\Lambda^{(4)}_{\overline{\mbox{\tiny MS}}}
=325\,\mbox{MeV}$.}
\newpage
\begin{eqnarray*}
\begin{array}{|c||c|c|c|c|c|c|}\hline
\mu&0.6\,\,\mbox{GeV}&0.7\,\,\mbox{GeV}&0.8\,\,\mbox{GeV}&
0.9\,\,\mbox{GeV}&1.0\,\,\mbox{GeV} \\ 
\hline\hline 
y_{1} & 0.0 & 0.0 & 0.0 & 0.0 & 0.0 \\[0.3mm]
y_{2} & 0.0 & 0.0 & 0.0 & 0.0 & 0.0 \\[0.3mm]
y_{3} & -0.020 & 0.039 & 0.043 & 0.041 & 0.039 \\[0.3mm]
y_{4} & -0.012 & -0.056 & -0.063 & -0.065 & -0.065 \\[0.3mm]
y_{5} & -0.447 & -0.092 & -0.036 & -0.017 & -0.008 \\[0.3mm]
y_{6} & -1.327 & -0.415 & -0.244 & -0.181 & -0.149 \\[0.3mm]
y_{7}/\alpha & -0.218 & -0.054 & -0.036 & -0.032 & -0.031 \\[0.3mm]
y_{8}/\alpha & 0.788 & 0.488 & 0.342 & 0.269 & 0.227 \\[0.3mm]
y_{9}/\alpha & -3.019 & -2.236 & -1.927 & -1.768 & -1.672 \\[0.3mm]
y_{10}/\alpha & 2.422 & 1.538 & 1.162 & 0.958 & 0.829 \\[0.3mm]
\hline
\end{array}
\end{eqnarray*}\\[2mm]\noindent
\centerline{Table~18:
$\Delta S = 1$ NLO Wilson coefficients (NDR) for
$\Lambda_{\mbox{\tiny QCD}}=\Lambda^{(4)}_{\overline{\mbox{\tiny MS}}}
=405\,\mbox{MeV}$.}
\vspace*{1.5cm}
\begin{eqnarray*}
\begin{array}{|c||c|c|c|c|c|c|}\hline
\mu&0.6\,\,\mbox{GeV}&0.7\,\,\mbox{GeV}&0.8\,\,\mbox{GeV}&
0.9\,\,\mbox{GeV}&1.0\,\,\mbox{GeV} \\ 
\hline\hline 
y_{1} & 0.0 & 0.0 & 0.0 & 0.0 & 0.0 \\[0.3mm]
y_{2} & 0.0 & 0.0 & 0.0 & 0.0 & 0.0 \\[0.3mm]
y_{3} & 0.023 & 0.059 & 0.055 & 0.050 & 0.046 \\[0.3mm]
y_{4} & -0.053 & -0.072 & -0.072 & -0.071 & -0.070 \\[0.3mm]
y_{5} & 0.217 & 0.048 & 0.026 & 0.021 & 0.019 \\[0.3mm]
y_{6} & -0.620 & -0.272 & -0.183 & -0.145 & -0.125 \\[0.3mm]
y_{7}/\alpha & 0.051 & -0.019 & -0.026 & -0.028 & -0.028 \\[0.3mm]
y_{8}/\alpha & 1.206 & 0.582 & 0.385 & 0.296 & 0.246 \\[0.3mm]
y_{9}/\alpha & -3.154 & -2.267 & -1.939 & -1.774 & -1.676 \\[0.3mm]
y_{10}/\alpha & 2.589 & 1.587 & 1.187 & 0.975 & 0.842 \\[0.3mm]
\hline
\end{array}
\end{eqnarray*}\\[2mm]\noindent
\centerline{Table~19:
$\Delta S = 1$ NLO Wilson coefficients (HV) for
$\Lambda_{\mbox{\tiny QCD}}=\Lambda^{(4)}_{\overline{\mbox{\tiny MS}}}
=405\,\mbox{MeV}$.}
\vspace*{\fill}
\end{appendix}
%
\newpage
\renewcommand{\textfraction}{1.0}
\end{document}